\documentclass[fleqn,3p,onecolumn,11pt,nofootinbib,showkeys,showpacs]{revtex4-1}

\usepackage[small,justification=centerlast]{caption}
\usepackage{graphicx}
\usepackage{bm}
\usepackage{mathrsfs}
\usepackage[latin1]{inputenc}
\usepackage{stmaryrd}
\usepackage{array}
\usepackage{psfrag}
\usepackage{dsfont}
\usepackage{epsfig}
\usepackage{titletoc}
\usepackage{float}   
\usepackage{wrapfig} 
\usepackage{amsmath}
\usepackage[latin1]{inputenc}
\usepackage{amsmath}\usepackage{amssymb,stmaryrd}
\usepackage{mathrsfs}
\usepackage{array}
\usepackage{graphicx}
\usepackage{psfrag}
\usepackage{dsfont}
\usepackage{epsfig}
\usepackage{cancel}
\usepackage[dvips]{feynmp}
\usepackage{upgreek}
\usepackage{subfig}
\usepackage{tabularx}
\usepackage{color}
\usepackage{units}

\usepackage{textfit} 

\usepackage{hyperref}

\renewcommand{\eqref}[1]{\mbox{Eq.~(\ref{#1})}}

\newcommand{\figref}[1]{\mbox{Fig.~\ref{#1}}}
\newcommand{\secref}[1]{\mbox{Sec.~\ref{#1}}}

\newcommand{\Secref}[1]{\mbox{Section \ref{#1}}}

\begin{document}

\title{Quantum field theory based on birefringent modified Maxwell theory}

\author{M. Schreck} \email{mschreck@indiana.edu}
\affiliation{Indiana University Center for Spacetime Symmetries, Indiana University, Bloomington, Indiana 47405-7105}

\begin{abstract}

In the current paper the properties of a birefringent Lorentz-violating extension of quantum electrodynamics is
considered. The theory results from coupling modified Maxwell theory, which is a \textit{CPT}-even Lorentz-violating
extension of the photon sector, to a Dirac theory of standard spin-1/2 particles. It is then restricted to a special
birefringent case with one nonzero Lorentz-violating coefficient.
The modified dispersion laws of electromagnetic waves are obtained plus their phase and group velocities are
considered. After deriving the photon propagator and the polarization vectors for a special momentum configuration
we prove both unitarity at tree-level and microcausality for the quantum field theory based on this Lorentz-violating
modification. These analytical proofs are done for a spatial momentum with two vanishing components and the proof of
unitarity is supported by numerical investigations in case all components are nonvanishing. The upshot is that the theory
is well-behaved within the framework of our assumptions where there is a possible issue for negative Lorentz-violating
coefficients. The paper shall provide a basis for the future analysis of alternative birefringent quantum field theories.

\end{abstract}

\maketitle

\newpage
\setcounter{equation}{0}
\setcounter{section}{0}
\renewcommand{\theequation}{\arabic{section}.\arabic{equation}}

\section{Introduction}

An optical medium is called birefringent if it has two different refraction indices depending on the corresponding polarization of
propagating light. Birefringence has various sources. There are solids that are birefringent without external manipulation since they
have one or several optical axes. The most prominent example for such a material is calcite. However, birefringence can also be caused
by external stress or electric and magnetic fields.

On the one hand, since the publication of the pioneering work by Klein and Nigam in 1964 it has been known that under certain circumstances
even the vacuum itself may become birefringent~\cite{Klein:1964zza}. Quantum field theory teaches us that the vacuum is not empty but filled
with a ``soup'' of virtual particles that are created and annihilated again. That is why the vacuum can be polarized by
applying an electric or magnetic field. Adler pointed out that peculiar effects occur for a photon traveling in strong magnetic
fields such as dispersion or splitting processes~\cite{Adler:1971wn}. The underlying reason for this behavior is that a strong external
field results in a nontrivial refraction index of the vacuum. The latter depends on the polarization state of the propagating photon, what
can then be interpreted as a birefringent vacuum \cite{Dittrich:1998gt,Battesti:2012hf}.
Experiments to detect this kind of vacuum birefringence include pulsed magnetic fields \cite{Bielsa:2009dz,Cadene:2013},
optical cavities with macroscopic magnetic fields \cite{Zavattini:2005ca,DellaValle:2013dwa}, and short pulses of conventional laser
systems \cite{Heinzl:2006xc}. New ideas even involve laser wake field acceleration \cite{Korn:2009} and free electron lasers \cite{Schlenvoigt:2013}.

On the other hand, a birefringent vacuum may occur in the context of quantum gravity. String theory
\cite{Kostelecky_Samuel:1989,Alan Kostelecky_Potting:1991,Alan Kostelecky_Potting:1996}, loop quantum gravity \cite{Gambini:1998it,Bojowald:2004bb},
theories of noncommutative spacetimes \cite{Carroll:2001ws}, and quantum field theories on spacetimes with nontrivial topologies
\cite{Klinkhamer:1998,Klinkhamer:2000,Klinkhamer:2003ec} give good arguments for a violation of Lorentz symmetry at the Planck scale.
Since there is still no established theory of quantum gravity available, we have
to rely on an effective framework that is suitable to describe Lorentz violation at energies much lower than the Planck energy. This
framework is called the Standard Model Extension (SME) \cite{ColladayKostelecky1998}. The SME extends the Lagrange density of the
Standard Model of elementary particle physics and of general relativity by all Lorentz-violating terms that respect the corresponding
gauge symmetries but violate particle Lorentz invariance. Each term is made up of Standard Model fields and Lorentz-violating
coefficients that determine the amount of Lorentz violation and that can be interpreted as fixed background fields. These fields
give rise to preferred directions in spacetime.

Some of the Lorentz-violating terms of the photon sector lead to a birefringent vacuum. It was observed that such terms either originate
from the breakdown of {\em CPT}-symmetry\footnote{Due to the {\em CPT} theorem, Lorentz invariance and {\em CPT} invariance are directly linked to
each other. However a violation of Lorentz invariance does not necessarily imply that {\em CPT} invariance is violated as well. For example,
the photon sector of the SME is made up of a {\em CPT}-odd and a {\em CPT}-even term. Recent results even indicate that a violation of {\em CPT}
invariance does not inevitably lead to Lorentz violation, e.g., in the context of noncommutative spacetimes
\cite{SheikhJabbari:2000vi,Chaichian:2002vw,Chaichian:2011fc,Chaichian:2012ga}.} or the dual symmetry of electrodynamics \cite{Kostelecky:2009zp}.
The first is associated
with Maxwell--Chern--Simons (MCS) theory \cite{Carroll_Field_Jackiw:1990}, whose Lagrange density combines a preferred spacetime
direction with the vector potential and the electromagnetic field strength tensor. The latter is connected to modified Maxwell theory \cite{ChadhaNielsen1983,ColladayKostelecky1998,KosteleckyMewes2002}, which involves a tensor-valued background field and a bilinear
combination of field strength tensors. The corresponding Lorentz-violating coefficients
can be bounded, e.g., by investigating the light of various astrophysical point sources or the cosmic microwave background itself.
This leads to very strict bounds in the order of $10^{-43}$ to $\unit[10^{-42}]{GeV}$ for the coefficients of MCS-theory and
$10^{-37}$ for the dimensionless coefficients of modified Maxwell theory (see \cite{Kostelecky:2008ts} and references therein).

Nevertheless, for theoretical reasons it is still interesting to gain a better understanding for the birefringent sectors of the
SME. This especially concerns the quantum field theories that are based on such a sector. MCS-theory has already been extensively
studied \cite{AdamKlinkhamer2001,Lehnert:2004hq,Lehnert:2004be,Lehnert:2005rh,Kaufhold_Klinkhamer:2006,Kaufhold:2007qd}. However
this is not the case for birefringent modified Maxwell theory. The goal of the current article is to extend our knowledge of these
special kinds of Lorentz-violating quantum field theories. The investigations performed can perhaps be adapted to other birefringent
quantum field theories as well. Recently, two articles were published where certain issues on the quantization procedure
according to Gupta and Bleuler were considered for birefringent modified Maxwell theory \cite{Colladay:2013dra,Colladay:2014dua}.

Note that birefringent photon dispersion relations also play a role in gravitational physics. In their seminal work from 1980,
Drummond and Hathrell showed that the classical trajectory of a photon in a gravitational background can be modified when quantum
effects are taken into account \cite{Drummond:1979pp}. This may lead to birefringence for certain curved spacetimes (see also the
review~\cite{Shore:2003zc} for a discussion of further questions on this issue).

The paper is organized as follows. In \secref{sec:birefringent-modmax} the action of modified Maxwell theory is introduced
and its most important properties are reviewed. The theory is then restricted to the birefringent sector with one nonzero coefficient.
The photon sector is coupled to a standard Dirac theory of spin-1/2 fermions to obtain
a birefringent extension of quantum electrodynamics (QED). \Secref{sec:dispersion-relations-classical-theory} is dedicated
to investigating the modified dispersion relations of the classical theory. Besides, both the phase and group velocities are
obtained and discussed. In \secref{sec:propagator-polarization-vectors} the propagator and the polarization vectors of the
corresponding quantum field theory are calculated that are used to investigate unitarity in \secref{sec:discussion-of-unitarity}
and microcausality \secref{sec:discussion-of-microcausality}. Finally the results are presented in the last section.
Calculational details are relegated to the appendix. Throughout the paper natural units are used with $\hbar=c=1$ unless stated otherwise.

\section{The birefringent sector of modified Maxwell theory}
\label{sec:birefringent-modmax}
\setcounter{equation}{0}

Modified Maxwell theory~\cite{ChadhaNielsen1983,ColladayKostelecky1998,KosteleckyMewes2002}, which is the \textit{CPT}-even modification
of the photon sector of the minimal SME,\footnote{Power-counting renormalizable Lorentz-violating terms are part of the ``minimal SME,''
whereas higher-dimensional terms form the ``nonminimal SME'' (see \cite{Kostelecky:2009zp,Kostelecky:2011gq,Kostelecky:2013rta} for the
nonminimal photon, neutrino, and fermion sector).} forms the basis of this paper. This theory is described by the following
action:\begin{subequations}\label{eq:action-modified-maxwell-theory}
\begin{eqnarray}
S_{\mathrm{modMax}}&=&\int_{\mathbb{R}^4}\mathrm{d}^4x\,
\mathcal{L}_\text{modMax}(x)\,,\\[2mm]
\label{eq:lagrange-density-modmax}
\mathcal{L}_\text{modMax}(x)&=& -\frac{1}{4}\,
\eta^{\mu\rho}\,\eta^{\nu\sigma}\,F_{\mu\nu}(x)F_{\rho\sigma}(x)
-\frac{1}{4}\,
\kappa^{\mu\nu\varrho\sigma}\,F_{\mu\nu}(x)F_{\varrho\sigma}(x)\,,
\label{eq:L-modified-maxwell-theory}
\end{eqnarray}
\end{subequations}
where $F_{\mu\nu}(x)\equiv\partial_{\mu}A_{\nu}(x)-\partial_{\nu}A_{\mu}(x)$ is the field strength tensor of the \textit{U}(1)
gauge field $A_{\mu}(x)$. All fields are defined on Minkowski spacetime with coordinates $(x^\mu)=(x^0,\boldsymbol{x})=(c\,t,x^1,x^2,x^3)$
and metric $(g_{\mu\nu}(x))=(\eta_{\mu\nu}) \equiv \text{diag}\,(1,-1,-1,-1)$. The first term on the right-hand side of the Lagrange density
of \eqref{eq:lagrange-density-modmax} is the standard Maxwell term and the second is the modification. The four-tensor
$\kappa^{\mu\nu\varrho\sigma}$
describes a background field, i.e., it transforms covariantly with respect to observer Lorentz transformations but it is fixed with
respect to particle Lorentz transformations. For this reason the second term violates particle Lorentz invariance.

The background tensor $\kappa^{\mu\nu\varrho\sigma}$ is antisymmetric under the interchange of the first two and the last two indices
plus it is symmetric when interchanging both index pairs:
\begin{subequations}
\label{eq:symmetries-background-tensor}
\begin{equation}
\kappa^{\mu\nu\varrho\sigma}=-\kappa^{\nu\mu\varrho\sigma}\,,\quad \kappa^{\mu\nu\varrho\sigma}=-\kappa^{\mu\nu\sigma\varrho}\,,\quad \kappa^{\mu\nu\varrho\sigma}=\kappa^{\varrho\sigma\mu\nu}\,.
\end{equation}
Furthermore it obeys the Bianchi identity
\begin{equation}
\sum_{(\nu,\varrho,\sigma)} \kappa^{\mu\nu\varrho\sigma}\equiv \kappa^{\mu\nu\varrho\sigma}+\kappa^{\mu\varrho\sigma\nu}+\kappa^{\mu\sigma\nu\varrho}=0\,,
\end{equation}
\end{subequations}
where the summation runs over cyclic permutations of the rightmost three indices of $\kappa^{\mu\nu\varrho\sigma}$. A nonvanishing double
trace $\kappa^{\mu\nu}_{\phantom{\mu\nu}\mu\nu}$ can be absorbed by a redefinition of the gauge field $A_{\mu}(x)$ and, therefore, it does
not describe any physics~\cite{ColladayKostelecky1998}. That is why it is usually set to zero: $\kappa^{\mu\nu}_{\phantom{\mu\nu}\mu\nu}=0$.
The above mentioned conditions reduce the number of independent coefficients of the background field to 19. Ten out of these 19 coefficients
lead to birefringent photon dispersion laws, i.e., there are two physical photon modes that have different phase velocities.

The remaining 9 coefficients result in photon dispersion relations that are nonbirefringent, at least at first order in the
Lorentz-violating coefficients. The nonbirefringent sector is parameterized by the following \textit{ansatz}~\cite{BaileyKostelecky2004,Altschul:2006zz}:
\begin{equation}\label{eq:nonbirefringent-Ansatz}
\kappa^{\mu\nu\varrho\sigma}=
\frac{1}{2}\,\Big(
 \eta^{\mu\varrho}\,\widetilde{\kappa}^{\nu\sigma}
-\eta^{\mu\sigma}\,\widetilde{\kappa}^{\nu\varrho}
-\eta^{\nu\varrho}\,\widetilde{\kappa}^{\mu\sigma}
+\eta^{\nu\sigma}\,\widetilde{\kappa}^{\mu\varrho}\Big)\,,
\end{equation}
where $\widetilde{\kappa}^{\mu\nu}$ is a symmetric and traceless $(4\times 4)$-matrix. In special cases the latter matrix
is usually constructed with two four-vectors $\xi^{\mu}$ and $\zeta^{\mu}$ (see, e.g., \cite{Kaufhold:2007qd,Casana-etal2010}) according to:
\begin{equation}
\label{eq:construction-symmetric-matrix}
\widetilde{\kappa}^{\mu\nu}=\frac{1}{2}(\xi^{\mu}\zeta^{\nu}+\zeta^{\mu}\xi^{\nu})-\frac{1}{4}(\xi\cdot \zeta)\,\eta^{\mu\nu}\,.
\end{equation}
The four-vectors $\xi^{\mu}$ and $\zeta^{\mu}$ are interpreted as preferred directions in spacetime. Hence, Lorentz violation in the
minimal sector of the SME leading to nonbirefringent photon dispersion relations is connected to having two preferred directions at the
maximum. For example, for the isotropic case of modified Maxwell theory one purely timelike four-vector is sufficient \cite{Klinkhamer:2010zs},
whereas for the parity-odd sector one purely timelike and one purely spacelike four-vector is needed \cite{Schreck:2011ai}.

In a series of papers it was demonstrated that a modified-Maxwell type term in the photon sector can arise in the one-loop effective
action of a modified QED \cite{Gomes:2009ch,Nascimento:2010cp,Scarpelli:2013eya}. The modification is constructed by a nonminimal
Lorentz-violating coupling between the photon and the fermion plus an axial Lorentz-violating term in the pure fermion sector.\footnote{
The latter contribution is the same that generates the four-dimensional \textit{CPT}-odd MCS-term via radiative corrections.} For
the nonminimal coupling term in four dimensions the effective action depends on the regularization procedure. Similar
issues arise in the context of the Adler--Bell--Jackiw anomaly, which is directly connected to the MCS-term in four dimensions. Hence
the papers show that such ambiguities occur in a wider class of quantum field theories, i.e., Lorentz-violating theories that lead to
modified-Maxwell type terms by radiative corrections.

The quantum-field-theoretic properties of the nonbirefringent part of modified Maxwell theory have already been studied extensively
in several papers \cite{Casana-etal2009,Casana-etal2010,Klinkhamer:2010zs,Schreck:2011ai}. Therefore, the current article aims at the
quantum field theory of the birefringent sector. The experimental sensitivity for the birefringent coefficients lies at $10^{-37}$
\cite{Kostelecky:2008ts}, which means that they are already tightly bounded. Nevertheless, a better understanding of the structure
of quantum field theories exhibiting birefringent particle dispersion laws is still of theoretical interest.

\subsection{Restriction to a particular birefringent theory}

The birefringent part of modified Maxwell theory can be parameterized by the ten coefficients given in Eq.~(8) of \cite{KosteleckyMewes2002}.
In what follows we choose the framework with the nonzero coefficient $\kappa^{0123}\in \mathbb{R}$ and all others, which are
not related to $\kappa^{0123}$ by symmetry arguments, set to zero. This special setup will
be investigated throughout the rest of the paper. Note that contrary to the nonbirefringent sector (cf. \eqref{eq:nonbirefringent-Ansatz})
the preferred spacetime directions playing a role for the birefringent case are not evident. To find them we try to construct our special
$\kappa^{\mu\nu\varrho\sigma}$ solely in terms of fixed four-vectors. Since the nonbirefringent \textit{ansatz} of \eqref{eq:nonbirefringent-Ansatz}
already respects the symmetries of the background tensor the following generalized \textit{ansatz} is used to match the Lorentz-violating
coefficients:
\begin{equation}
\kappa^{\mu\nu\varrho\sigma}\Big|^{\begin{subarray}{l}
\mathrm{birefringent} \\
\mathrm{sector}
\end{subarray}}=\frac{\widetilde{\kappa}}{2}\,\Big(\widetilde{\kappa}_1^{\mu\varrho}\widetilde{\kappa}_2^{\nu\sigma}-\widetilde{\kappa}_1^{\mu\sigma}\widetilde{\kappa}_2^{\nu\varrho}-\widetilde{\kappa}_1^{\nu\varrho}\widetilde{\kappa}_2^{\mu\sigma}+\widetilde{\kappa}_1^{\nu\sigma}\widetilde{\kappa}_2^{\mu\varrho}\Big)\,,
\end{equation}
with the two symmetric and traceless $(4\times 4)$-matrices $\widetilde{\kappa}_1^{\mu\nu}$, $\widetilde{\kappa}_2^{\mu\nu}$ and the scalar
quantity $\widetilde{\kappa}$. The matrices are expressed in terms of two four-vectors analogously to \eqref{eq:construction-symmetric-matrix},
i.e., there is a total number of four such vectors denoted as $\xi_i$ for $i\in \{0,1,2,3\}$. Each matrix can contain only one or two
different vectors. For example the first is given by
\begin{equation}
\widetilde{\kappa}_1^{\mu\nu}=\widetilde{\kappa}_1^{\mu\nu}(\xi_i,\xi_j)=\frac{1}{2}(\xi_i^{\mu}\xi_j^{\nu}+\xi_j^{\mu}\xi_i^{\nu})-\frac{1}{4}(\xi_i\cdot \xi_j)\,\eta^{\mu\nu}\,.
\end{equation}
Here the indices $i$, $j\in \{0,1,2,3\}$ may but need not necessarily be equal. We choose the following set of orthonormal
four-vectors:
\begin{equation}
\label{eq:orthonormal-set-preferred-vectors}
\xi_0=\begin{pmatrix}
1 \\
0 \\
0 \\
0 \\
\end{pmatrix}\,,\quad \xi_1=\begin{pmatrix}
0 \\
1 \\
0 \\
0 \\
\end{pmatrix}\,,\quad \xi_2=\begin{pmatrix}
0 \\
0 \\
1 \\
0 \\
\end{pmatrix}\,,\quad \xi_3=\begin{pmatrix}
0 \\
0 \\
0 \\
1 \\
\end{pmatrix}\,,
\end{equation}
where the first is purely timelike and the others are purely spacelike. The aforementioned \textit{ansatz} is generic enough
to match the particular $\kappa^{\mu\nu\varrho\sigma}$ considered. This is done for all combinations of $\xi_i$ and leads to the following
result:
\begin{subequations}
\label{eq:generalized-ansatz-special-case}
\begin{align}
\kappa^{\mu\nu\varrho\sigma}\Big|^{\begin{subarray}{l}
\mathrm{birefringent} \\
\mathrm{sector}
\end{subarray}}_{\kappa^{0123}\neq 0}&=4\kappa^{0123}\,\Big(\widetilde{\kappa}_1^{\mu\varrho}\widetilde{\kappa}_2^{\nu\sigma}-\widetilde{\kappa}_1^{\mu\sigma}\widetilde{\kappa}_2^{\nu\varrho}-\widetilde{\kappa}_1^{\nu\varrho}\widetilde{\kappa}_2^{\mu\sigma}+\widetilde{\kappa}_1^{\nu\sigma}\widetilde{\kappa}_2^{\mu\varrho}\Big)\,, \\
\widetilde{\kappa}_1^{\mu\nu}&=\widetilde{\kappa}_1^{\mu\nu}(\xi_0,\xi_2)\,,\quad \widetilde{\kappa}_2^{\mu\nu}=\widetilde{\kappa}_2^{\mu\nu}(\xi_1,\xi_3)\,.
\end{align}
\end{subequations}
Consequently, the particular background tensor considered can be decomposed into one purely timelike and three purely spacelike
four-vectors according to \eqref{eq:generalized-ansatz-special-case}. This demonstration was made because of two
reasons. First, it was shown that the well-known nonbirefringent \textit{ansatz} of \eqref{eq:nonbirefringent-Ansatz} can be generalized
for the birefringent sector of modified Maxwell theory. Therefore, the nonbirefringend and birefringend sector may have a common
underlying structure. Second, with the result obtained it becomes clear that the special part of the birefringent sector, which forms the
basis of the paper, is characterized by the single nonvanishing coefficient $\kappa^{0123}$ and the orthogonal set of four preferred
spacetime directions given by \eqref{eq:orthonormal-set-preferred-vectors}. This finding will be used frequently throughout the paper.

\subsection{Coupling to matter: Birefringent extension of quantum electrodynamics}
\label{sec:coupling-to-standard-dirac-particles}

To construct a QED extension, the photon sector is coupled to standard spin-1/2 Dirac fermions with electric charge $e$ and mass $m$.
This is performed by employing the usual minimal coupling procedure. It leads to a birefringent modification of QED with the
following action:
\begin{equation}\label{eq:action-birefringent-qed} \hspace*{0mm}
S_\text{modQED}^\text{birefringent}\big[\kappa^{0123},e,m\big] =
S_\text{modMax}^\text{birefringent}\big[\kappa^{0123}\big] +
S^\text{}_\text{Dirac}\big[e,m\big]\,.
\end{equation}
The modified-Maxwell term for the gauge field $A_\mu(x)$ is given by Eqs. (\ref{eq:action-modified-maxwell-theory}),
(\ref{eq:generalized-ansatz-special-case}) and the standard Dirac term for the spinor field $\psi(x)$ is
\begin{subequations}
\label{eq:standDirac-action}
\begin{align}
S^\text{ }_\text{Dirac}\big[e,m\big] &=
\int_{\mathbb{R}^4} \mathrm{d}^4 x \; \overline\psi(x) \left[
\gamma^\mu \left(\frac{\mathrm{i}}{2}\,\overleftrightarrow{\partial_\mu} -e A_\mu(x) \right) -m\right] \psi(x)\,, \\[2ex]
A\overleftrightarrow{\partial_{\mu}}B&\equiv A\partial_{\mu}B-(\partial_{\mu}A)B\,.
\end{align}
\end{subequations}
Equation (\ref{eq:standDirac-action}) involves the standard Dirac matrices $\gamma^\mu$ obeying the Clifford algebra
$\{\gamma^{\mu},\gamma^{\nu}\}=2\eta^{\mu\nu}$. In the written form the Lagrange density is
granted to be Hermitian.

\section{Dispersion relations}
\label{sec:dispersion-relations-classical-theory}
\setcounter{equation}{0}

The field equations \cite{ColladayKostelecky1998,KosteleckyMewes2002,BaileyKostelecky2004} of modified Maxwell theory in momentum space
are given by:
\begin{equation}
\label{eq:field-equations-modified-maxwell-theory}
M^{\mu\nu}A_{\nu}=0 \,,\quad
M^{\mu\nu}\equiv
k^{\rho}k_{\rho}\,\eta^{\mu\nu}-k^{\mu}k^{\nu}
-2\,\kappa^{\mu\rho\sigma\nu}\,k_{\rho}k_{\sigma}\,,
\end{equation}
where $k^{\mu}$ is the four-momentum.
Choosing a particular gauge fixing such as Lorenz gauge $k^{\mu}A_{\mu}=0$, the dispersion relations result from the condition
$\det(M)=0$ with the matrix $M$ given in \eqref{eq:field-equations-modified-maxwell-theory}. The dispersion relations of the
unphysical scalar and longitudinal modes can be identified using the modified Coulomb and Amp\`{e}re law (cf. the procedure
in \cite{ColladayKostelecky1998}). These are given by
\begin{equation}
\label{eq:dispersion-relations-unphysical}
\omega_0(\mathbf{k})=\omega_3(\mathbf{k})=|\mathbf{k}|\,,
\end{equation}
where the scalar mode is marked by the index $\lambda=0$ and the longitudinal mode by the index $\lambda=3$.
This is in concordance with the investigations performed for the nonbirefringent sector \cite{Klinkhamer:2010zs,Schreck:2011ai}. The
unphysical dispersion laws always correspond to the standard relations, i.e., they are unaffected by Lorentz violation, which directly
follows from the modified field equations \cite{Colladay:2014dua}. On the
contrary, the dispersion relations for the two physical degrees of freedom of electromagnetic waves (labeled by indices $\lambda=1$, 2)
are heavily modified and given by the following result:
\begin{subequations}
\label{eq:dispersion-relations-birefringent}
\begin{equation}
\omega_{1,2}(\mathbf{k})=\frac{1}{\sqrt{3}}\left(f(\mathcal{C}_1,\mathcal{C}_2,\mathcal{C}_5)\pm \sqrt{3\mathcal{C}_5-f(\mathcal{C}_1,\mathcal{C}_2,\mathcal{C}_5)^2-\frac{24\sqrt{3}\,k_1k_2k_3\,(\kappa^{0123})^3}{f(\mathcal{C}_1,\mathcal{C}_2,\mathcal{C}_5)}}\,\right)\,,
\end{equation}
\begin{equation}
f(\mathcal{C}_1,\mathcal{C}_2,\mathcal{C}_5)=\sqrt{\sqrt[3]{\mathcal{C}_1}+\frac{\mathcal{C}_2}{\sqrt[3]{\mathcal{C}_1}}+\mathcal{C}_5}\,,
\end{equation}
\begin{align}
\mathcal{C}_1&=\frac{1}{128}\left[\mathcal{C}_3+\sqrt{\mathcal{C}_3^2-256(\mathcal{C}_5^2+3\mathcal{C}_4)^3}\right]\,,\quad \mathcal{C}_2=\frac{1}{4}(\mathcal{C}_5^2+3\mathcal{C}_4)\,, \\[2ex]
\mathcal{C}_3&=16\left[9\mathcal{C}_5\mathcal{C}_4-\mathcal{C}_5^3+1728k_1^2k_2^2k_3^2\,(\kappa^{0123})^6\right]\,, \\[2ex]
\mathcal{C}_4&=\mathbf{k}^4+4(k_1^2k_2^2+4k_1^2k_3^2+k_2^2k_3^2)(\kappa^{0123})^2\,,\quad \mathcal{C}_5=\mathbf{k}^2+2(\mathbf{k}^2+3k_2^2)(\kappa^{0123})^2\,,
\end{align}
\end{subequations}
for a general three-momentum $\mathbf{k}=(k_1,k_2,k_3)$. An expansion for $|\kappa^{0123}|\ll 1$ to linear order yields an approximation,
which coincides with Eq.~(16) in \cite{KosteleckyMewes2002}:
\begin{equation}
\label{eq:dispersion-relation-expanded}
\omega_{1,2}(\mathbf{k})=|\mathbf{k}|\pm \frac{\sqrt{(k_1^2+2k_2^2+k_3^2)^2-4k_1^2k_3^2}}{|\mathbf{k}|}\,\kappa^{0123}+\mathcal{O}[(\kappa^{0123})^2]\,.
\end{equation}
The radicand is nonnegative for all choices of the three-momentum components. Hence, the dispersion relations are real at first order Lorentz
violation. Whether or not this is the case for the exact expressions is currently not clear because of their complexity. Note that, e.g.,
for isotropic modified Maxwell theory or special anisotropic sectors the exact dispersion relations are not necessarily real for all values
of Lorentz-violating coefficients or three-momentum components \cite{Klinkhamer:2010zs}. Besides, for certain three-momenta, for example
$\mathbf{k}=(k,0,k)$, the first order term in \eqref{eq:dispersion-relation-expanded} even vanishes. For such propagation directions the
theory is birefringent only at higher orders Lorentz violation.

Equation (30) in \cite{Kostelecky:2009zp} shows that the off-shell dispersion relation of modified Maxwell theory (i.e., the equation whose
zeros with respect to $k^0$ correspond to the dispersion relations) can be written in a very compact form. For this purpose the first of the
following definitions is used and we add a second convenient definition:
\begin{equation}
(\widetilde{\wedge^2M})_{\mu\alpha\nu\beta}\equiv \frac{1}{4}\varepsilon_{\mu\alpha\varrho\gamma}\varepsilon_{\nu\beta\sigma\delta}M^{\varrho\sigma}M^{\gamma\delta}\,,\quad
(\widetilde{\wedge^2M_{\kappa}})_{\mu\alpha\nu\beta}\equiv \frac{1}{4}\varepsilon_{\mu\alpha\varrho\gamma}\varepsilon_{\nu\beta\sigma\delta}M_{\kappa}^{\varrho\sigma}M_{\kappa}^{\gamma\delta}\,,
\end{equation}
Herein, $M$ is given in \eqref{eq:field-equations-modified-maxwell-theory} and $M_{\kappa}$ is its Lorentz-violating part:
$M_{\kappa}^{\mu\nu}\equiv -2\,\kappa^{\mu\rho\sigma\nu}\,k_{\rho}k_{\sigma}$.
We hereby stick to the notation that is used in the latter reference. The wedge ``$\wedge$'' denotes the outer product of two four-vectors,
which is widely employed in \cite{Kostelecky:2009zp}. Now the following relationships hold:
\begin{subequations}
\begin{align}
\frac{1}{k^0}\kappa^{\mu\alpha\nu\beta}(\widetilde{\wedge^2M_{\kappa}})_{\mu\alpha\nu\beta}&=-96k_1k_2k_3(\kappa^{0123})^3\,, \\[2ex]
\Xi\big|_{k^0=0}&=4(k_1^2k_2^2+4k_1^2k_3^2+k_2^2k_3^2)(\kappa^{0123})^2\,, \\[2ex]
\frac{1}{2(k^0)^2}\left[\Xi\big|_{k^0=0}-\Xi\right]&=2(\mathbf{k}^2+3k_2^2)(\kappa^{0123})^2\,,\quad \Xi\equiv \eta^{\mu\beta}\eta^{\nu\alpha}(\widetilde{\wedge^2M})_{\mu\alpha\nu\beta}-3k^4\,.
\end{align}
\end{subequations}
Comparing these results to \eqref{eq:dispersion-relations-birefringent}, it becomes clear that the respective terms in the physical
dispersion laws depending on the three-momentum components can, in principle, be written in a compact form as well. However the
explicit expressions are more suitable for the calculations that follow.

\begin{figure}[t]
\subfloat[]{\label{fig:plot-dispersion-1}\includegraphics[scale=0.75]{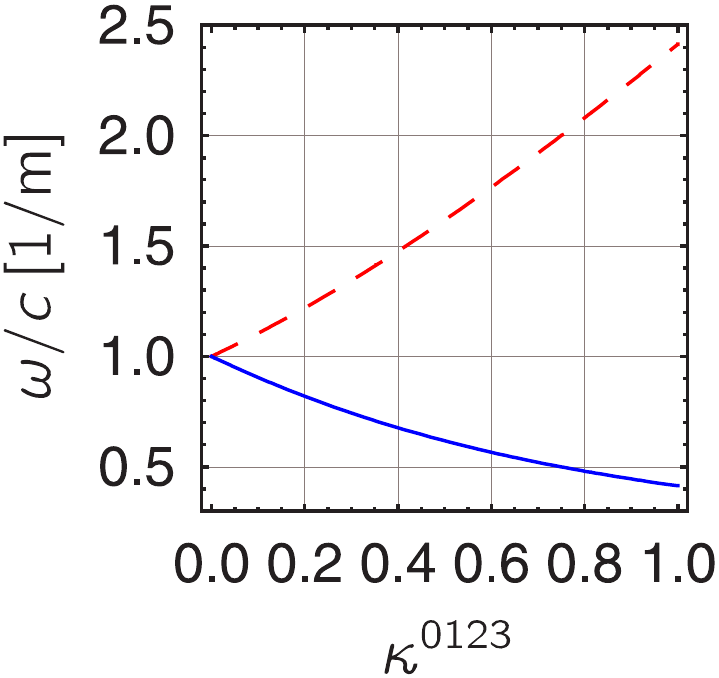}} \hspace{1cm}
\subfloat[]{\label{fig:plot-dispersion-2}\includegraphics[scale=0.75]{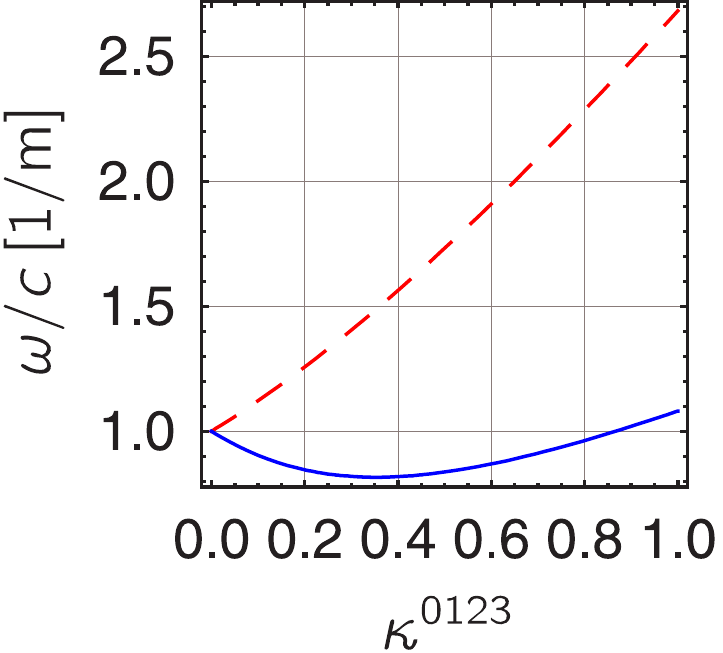}}
\caption{The current figures show the dispersion relations $\omega_{1,2}$ of \eqref{eq:dispersion-relations-birefringent} as a function
of $\kappa^{0123}$ for different values of the three-momentum components. The red (dashed) curves represent $\omega_1$ and the blue (plain)
curves correspond to~$\omega_2$. The values $k_1=k_2=0$, $k_3=\mathrm{1/m}$ were chosen in (a) and $k_1=k_2=k_3=\mathrm{1/(\sqrt{3}\,m)}$
was used in (b).}
\label{fig:plot-dispersion-relations}
\end{figure}

Both dispersion relations are shown in \figref{fig:plot-dispersion-relations} as functions of $\kappa^{0123}$. On the one hand, the
first dispersion law $\omega_1$ seems to increase monotonically for any three-momentum. On the other hand, the three-momentum can be
chosen such that $\omega_2$ decreases monotonically (see \figref{fig:plot-dispersion-1}). However there also exist choices for
which $\omega_2$ decreases until it reaches some minimum. After reaching the minimum it starts increasing again (see \figref{fig:plot-dispersion-2}).

\begin{figure}[t]
\subfloat[]{\label{fig:plot-nullcones-1}\includegraphics{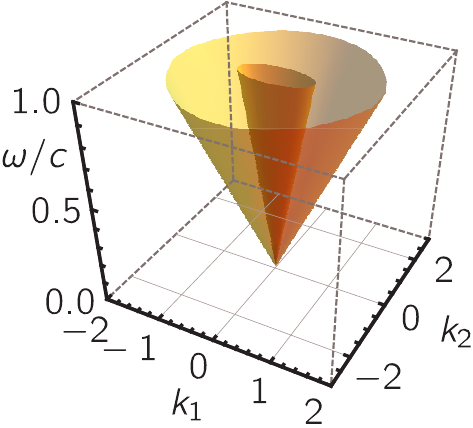}}\hspace{0.4cm}
\subfloat[]{\label{fig:plot-nullcones-2}\includegraphics{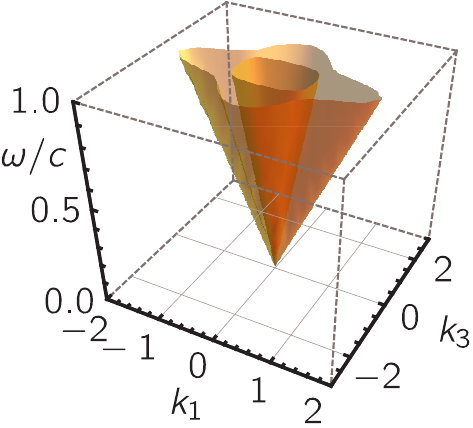}}\hspace{0.4cm}
\subfloat[]{\label{fig:plot-nullcones-3}\includegraphics{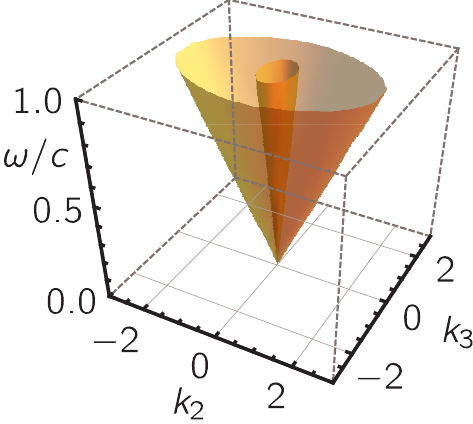}} \\
\subfloat[]{\label{fig:plot-nullcones-4}\includegraphics{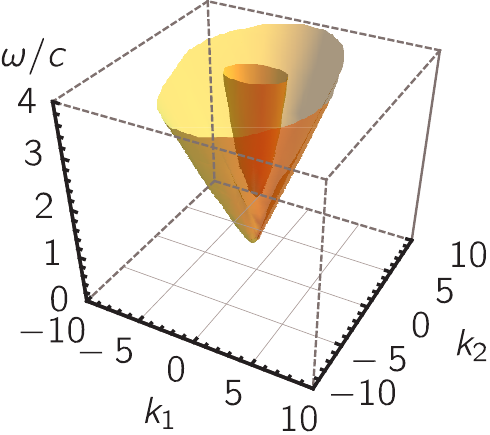}}\hspace{0.4cm}
\subfloat[]{\label{fig:plot-nullcones-5}\includegraphics{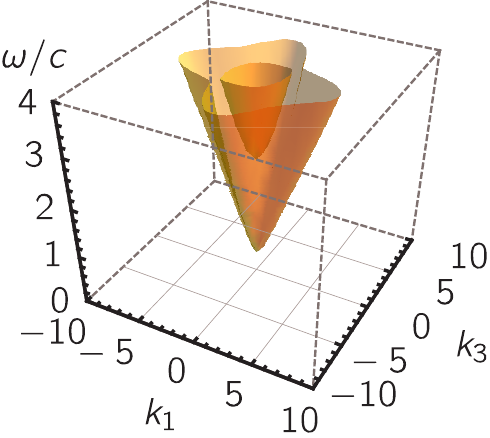}}\hspace{0.4cm}
\subfloat[]{\label{fig:plot-nullcones-6}\includegraphics{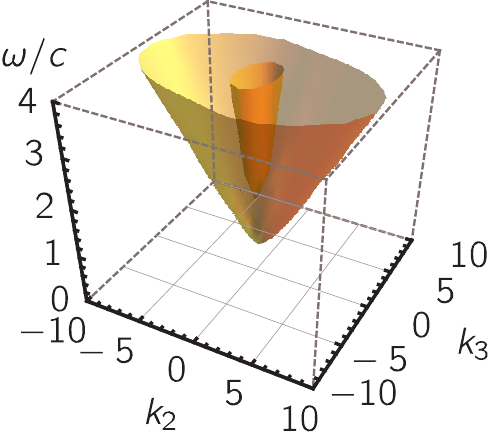}}
\caption{Two-dimensional slices of the modified photon dispersion relations given by \eqref{eq:dispersion-relations-birefringent}.
The surface corresponding to $\omega_1$ always lies in the interior of the surface that corresponds to $\omega_2$. To visualize
the distortion of the surfaces by Lorentz violation, the rather large value $\kappa^{0123}=1/2$ was chosen. Furthermore, the
fixed components of the spatial momentum are chosen as follows: $k_3=0$ (a), $k_2=0$ (b), $k_1=0$ (c), $k_3=\mathrm{1/m}$ (d),
$k_2=\mathrm{1/m}$ (e), and $k_1=\mathrm{1/m}$ (f).}
\label{fig:plot-nullcones}
\end{figure}%
In \figref{fig:plot-nullcones} two-dimensional slices of the physical dispersion relations are shown. The upper panels,
Figs.~\ref{fig:plot-nullcones-1} -- \ref{fig:plot-nullcones-3}, are obtained by setting one particular momentum component to zero for
each of the figures. The respective surfaces are distortions of the standard nullcones in momentum space. First, due to Lorentz violation
their opening angles change and second, intersections with planes parallel to the $k_i$-$k_j$-plane (where $k_i$, $k_j$ are the remaining
momentum components not set to a particular value) are no circles any more.

In the lower panels, Figs.~\ref{fig:plot-nullcones-4} -- \ref{fig:plot-nullcones-6}, the respective components are set to a nonzero
value. These surfaces are modifications of the standard hyperbola in momentum space. Looking at the figures it becomes
evident that the surfaces shown do not intersect each other. Hence there are no degeneracies in the range of momenta that is presented.
Whether or not this holds in general is an open problem, which will not be considered further.

\begin{figure}[t]
\centering
\includegraphics[scale=0.75]{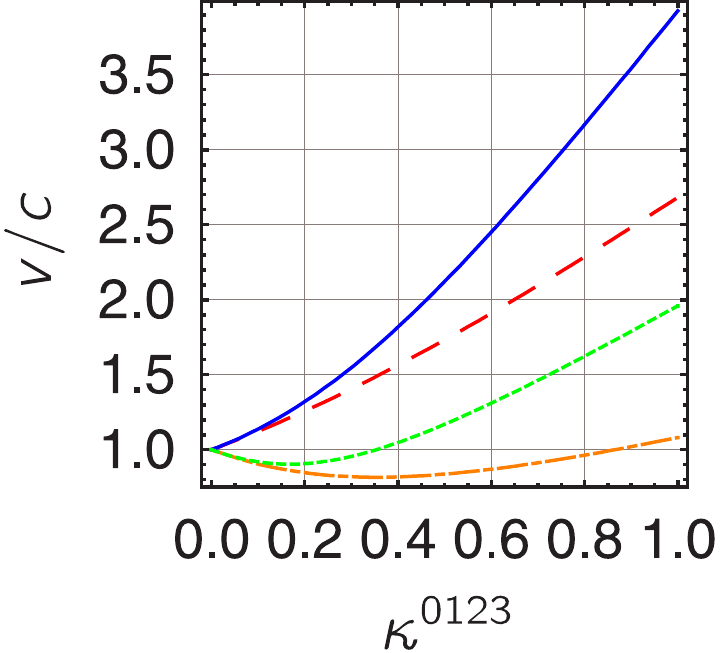}
\caption{Phase and group velocities as a function of $\kappa^{0123}$ for a special three-momentum with $k_1=k_2=k_3=\mathrm{1/(\sqrt{3}\,m)}$.
The phase velocities of the first/second mode are shown as a red/orange (dashed/dashed-dotted) curve. The respective group velocities are
drawn in blue (plain) and green (dotted), respectively.}
\label{fig:wave-velocities}
\end{figure}

As a next step we are interested in the phase and group velocities of classical electromagnetic waves. These are given as follows:
\begin{equation}
v_{\mathrm{ph};1,2}=\frac{\omega_{1,2}(\mathbf{k})}{|\mathbf{k}|}\,,\quad v_{\mathrm{gr};1,2}=\left|\frac{\partial\omega_{1,2}(\mathbf{k})}{\partial\mathbf{k}}\right|\,.
\end{equation}
Our interest especially concerns the values of the phase velocity for infinite three-momentum. They are connected to the front
velocity being defined as \cite{Brillouin1960}
\begin{equation}
v_{\mathrm{fr};1,2}\equiv \lim_{|\mathbf{k}|\mapsto \infty} v_{\mathrm{ph};1,2}\,,
\end{equation}
and correspond to the propagation velocity of a $\delta$-function shaped signal. Hence the front velocity is associated with information
transport of a wave. The exact expressions are involved and there is not much insight to be gained from them. Therefore, we give expansions in
the Lorentz-violating coefficient $\kappa^{0123}$. Since the theory is anisotropic, the results depend on the propagation direction considered.
As examples we choose the $x$-, $y$- and $z$-direction. For the first physical photon mode the results are:
\begin{subequations}
\begin{align}
\lim_{k_1\mapsto \infty} v_{\mathrm{ph},1}&=\lim_{k_3\mapsto \infty} v_{\mathrm{ph},1}=1+\kappa^{0123}+\frac{1}{2}(\kappa^{0123})^2-\frac{1}{8}(\kappa^{0123})^4\pm \dots\,, \\[2ex]
\lim_{k_2\mapsto \infty} v_{\mathrm{ph},1}&=1+2\kappa^{0123}+2(\kappa^{0123})^2-2(\kappa^{0123})^4\pm \dots\,.
\end{align}
\end{subequations}
For the second photon mode we obtain:
\begin{subequations}
\begin{align}
\lim_{k_1\mapsto \infty} v_{\mathrm{ph},2}&=\lim_{k_3\mapsto \infty} v_{\mathrm{ph},2}=1-\kappa^{0123}+\frac{1}{2}(\kappa^{0123})^2-\frac{1}{8}(\kappa^{0123})^4\pm \dots\,, \\[2ex]
\lim_{k_2\mapsto \infty} v_{\mathrm{ph},2}&=1-2\kappa^{0123}+2(\kappa^{0123})^2-2(\kappa^{0123})^4\pm \dots\,.
\end{align}
\end{subequations}
In the limit considered the group velocities correspond to the phase velocities. It is evident that for $|\kappa^{0123}|\ll 1$ the first
photon mode is superluminal whereas the second is subluminal.\footnote{Note that the maximum attainable velocity of standard Dirac particles
is given by $c$, which we set equal to 1 throughout the paper. By ``sublimunal'' we mean that a wave propagates with a velocity smaller
than $c$. If we use the term ``superluminal'' it travels faster than $c$.} The respective expansions of $\omega_1$ und $\omega_2$ correspond
to each other except the linear term in $\kappa^{0123}$, which comes with a different sign. Hence this linear term is what determines whether
the wave travels faster or slower than the maximum attainable velocity $c$ of standard Dirac particles. This also reflects the birefringent
properties of the theory considered. See \figref{fig:wave-velocities} for a plot of the phase and group velocities for a special
three-momentum chosen.

In the literature it is sometimes conjectured that superluminal modes lead to problems with (micro)causality. However various studies
have shown that this is not an issue at all for a wide range of Lorentz-violating frameworks (see, e.g.,
\cite{KosteleckyLehnert2000,AdamKlinkhamer2001,Liberati:2001sd,Klinkhamer:2010zs,Schreck:2011ai}).\footnote{Note that superluminal modes also
appear in certain cosmological frameworks. For example, in \cite{Babichev:2007dw} it was proven that superluminal front velocities do not lead
to problems in k-essence models.} The important point in this context
can be summarized with few sentences. Microcausality for a electromagnetic theory is guaranteed as long as information propagates along
or inside nullcones \cite{Dubovsky:2007ac}. Whether or not these nullcones are modified by Lorentz violation does not change this argument.

\subsection{Special limit of the dispersion relations}
\label{sec:limit-dispersion-relation}

For a photon propagating with the three-momentum $\mathbf{k}=(0,k_2,k_3)$ where $k_3>0$ an interesting
observation can be made. Its (squared) dispersion relations, which follow from \eqref{eq:field-equations-modified-maxwell-theory}, are
given by:
\begin{align}
\omega_{1,2}(0,k_2,k_3)^2&=k_2^2+k_3^2+2(4k_2^2+k_3^2)(\kappa^{0123})^2 \notag \\
&\phantom{{}={}}\pm 2|\kappa^{0123}|\sqrt{(2k_2^2+k_3^2)^2+(4k_2^2+k_3^2)^2(\kappa^{0123})^2}\,.
\end{align}
The result was squared just for the matter of conveniently writing the formulas. In the limit $k_2\mapsto 0$ the physical
dispersion laws result in:
\begin{equation}
\omega_{1,2}(0,0,k_3)=k_3\sqrt{1+2(\kappa^{0123})^2\pm 2|\kappa^{0123}|\sqrt{1+(\kappa^{0123})^2}}\,.
\end{equation}
This can be rewritten as follows:
\begin{equation}
\label{eq:dispersion-relation-special-case-2}
\omega_{1,2}(0,0,k_3)=k_3\sqrt{\left[\sqrt{1+(\kappa^{0123})^2}\pm |\kappa^{0123}|\right]^2}=k_3\left[\sqrt{1+(\kappa^{0123})^2}\pm |\kappa^{0123}|\right]\,.
\end{equation}
Now this result should correspond to the modified photon dispersion relations that can be obtained directly from
\eqref{eq:field-equations-modified-maxwell-theory} by setting $k_1=k_2=0$ and assuming $k_3>0$ at the beginning:
\begin{equation}
\label{eq:dispersion-relation-special-case-k3positive}
\omega_{1,2}(k_3)=k_3\left[\sqrt{1+(\kappa^{0123})^2}\pm \kappa^{0123}\right]\,.
\end{equation}
Note that the sign of a nonzero $\kappa^{0123}$ is not determined by the theory but the decision is taken by nature. Hence $\kappa^{0123}$
can be positive, zero or negative. In the second term on the far right-hand side of \eqref{eq:dispersion-relation-special-case-2} there appears
the absolute value of $\kappa^{0123}$, which is not the case in \eqref{eq:dispersion-relation-special-case-k3positive}. Hence for nonnegative $\kappa^{0123}$
the limit obtained from the general dispersion laws corresponds to the actual dispersion laws of a photon traveling with a three-momentum
$\mathbf{k}=(0,0,|\mathbf{k}|)=(0,0,k_3)$, which is given by \eqref{eq:dispersion-relation-special-case-k3positive}. However this is not the
case for negative $\kappa^{0123}$.

This behavior originates from the general form of the dispersion laws of \eqref{eq:dispersion-relations-birefringent} being characterized by
the square root of expressions including further
square roots. Note that within certain Finsler structures double square roots appear as well (see, e.g.,
Eq.~(17) in \cite{Kostelecky:2010hs}).
They are connected to degeneracies of the corresponding dispersion relations \cite{Kostelecky:2011qz}. From a mathematical point of view
energy-momentum space is then not a manifold any more but an algebraic variety~\cite{Kostelecky:2010hs}. Contrary to a manifold, a variety is
allowed to have singularities, i.e., points where no tangent vectors exist. If dispersion relations of different modes are, indeed, degenerate
in certain regions these may be singular in this sense. In 1964 it was proven by Hironaka that such singularities can be removed in principle
\cite{Hironaka:1964} where the detailed procedure depends on the variety to be studied. A better understanding of this issue in the context of
the Lorentz-violating photon theory considered is an interesting open problem to investigate further.

\section{Propagator and polarization vectors}
\label{sec:propagator-polarization-vectors}

The previous sections were dedicated to investigating the properties of the classical theory, especially the modified photon dispersion
laws. The most important result was that there are two propagation modes whose propagation velocities differ at first order Lorentz
violation, clearly indicating birefringence. For infinite momentum and $|\kappa^{0123}|\ll 1$ one of these modes travels slower than the
maximum velocity of standard Dirac particles and the other travels faster.

In the following sections we are interested in understanding the corresponding quantum theory. For this purpose we will obtain the gauge
propagator and the polarization vectors of the physical modes. The propagator is the Green's function of the free field equations
(\ref{eq:field-equations-modified-maxwell-theory}) in momentum space. It results from inverting the matrix $M$ of the latter equation
and expressing it in a covariant form. To do so, the gauge must be fixed because otherwise $M$ does not have an inverse due to the
infinite number of gauge degrees of freedom. The calculation will be performed using Feynman gauge
\cite{Veltman1994,ItzyksonZuber1980,PeskinSchroeder1995}, i.e., by adding the following gauge-fixing term to the Lagrange density:
\begin{equation}
\label{eq:gauge-fixing-feynman}
\mathcal{L}_{\mathrm{gf}}(x)=
-\frac{1}{2}\big(\partial_{\mu}\,A^{\mu}(x)\big)^2\,.
\end{equation}
Now we need a covariant \textit{ansatz} for the propagator $\widehat{G}^{\mu\nu}$. It is set up by using all two-rank tensors that play
a role in this theory. This is the metric and additional tensors that are constructed with the four-momentum and the preferred directions
given by \eqref{eq:orthonormal-set-preferred-vectors}. So it is given by:
\begin{align}\label{eq:propagator-parity-odd-coeff}
\widehat{G}^{\mu\nu}\,
\big|^{\mathrm{Feynman}}
=-\mathrm{i}\,\Big\{
&+\widehat{a}\,\eta^{\mu\nu}+\widehat{b}\,\xi_0^{\mu}\xi_0^{\nu}+\widehat{c}\,\xi_1^{\mu}\xi_1^{\nu}+\widehat{d}\,\xi_2^{\mu}\xi_2^{\nu}+\widehat{e}\,\xi_3^{\mu}\xi_3^{\nu} \notag \\
&+\widehat{f}(\xi_0^{\mu}\xi_1^{\nu}+\xi_1^{\mu}\xi_0^{\nu})+\widehat{g}(\xi_0^{\mu}\xi_2^{\nu}+\xi_2^{\mu}\xi_0^{\nu})+\widehat{h}(\xi_0^{\mu}\xi_3^{\nu}+\xi_3^{\mu}\xi_0^{\nu}) \notag \\
&+\widehat{i}(\xi_1^{\mu}\xi_2^{\nu}+\xi_2^{\mu}\xi_1^{\nu})+\widehat{j}(\xi_1^{\mu}\xi_3^{\nu}+\xi_3^{\mu}\xi_1^{\nu})+\widehat{k}(\xi_2^{\mu}\xi_3^{\nu}+\xi_3^{\mu}\xi_2^{\nu}) \notag \\
&+\widehat{l}\,k^{\mu}k^{\nu}+\widehat{m}\,(k^{\mu}\xi_0^{\nu}+\xi_0^{\mu}k^{\nu})+\widehat{n}(k^{\mu}\xi_1^{\nu}+\xi_1^{\mu}k^{\nu}) \notag \\
&+\widehat{o}(k^{\mu}\xi_2^{\nu}+\xi_2^{\mu}k^{\nu})+\widehat{p}(k^{\mu}\xi_3^{\nu}+\xi_3^{\mu}k^{\nu})\Big\}\,\widehat{K}\,.
\end{align}
The propagator coefficients are functions of the four-momentum components, i.e., $\widehat{a}=\widehat{a}(k^0,\mathbf{k})$,
\dots, $\widehat{p}=\widehat{p}(k^0,\mathbf{k})$. Moreover, there is a scalar part $\widehat{K}=\widehat{K}(k^0,\mathbf{k})$ that
can be factored out. The coefficients $\widehat{l}$, \dots, $\widehat{p}$ depend on
the gauge and the associated terms vanish when the propagator is contracted with a gauge-invariant quantity. This is due to the
Ward identity, which is still valid, since the QED extension is gauge-invariant and no anomalies are expected to occur.

All coefficients follow from the inversion of the matrix $M$ of \eqref{eq:field-equations-modified-maxwell-theory}. Hence the system $(\widehat{G}^{-1})^{\mu\nu}\widehat{G}_{\nu\lambda}=\mathrm{i}\,\delta^{\mu}_{\phantom{\mu}\lambda}$
has to be solved. Herein, $(\widehat{G}^{-1})^{\mu\nu}$ corresponds to the differential operator
\begin{equation}
\label{eq:differential-operator}
(G^{-1})^{\mu\nu}=\eta^{\mu\nu}\partial^2
-2\, \kappa^{\mu\varrho\sigma\nu}\partial_{\varrho}\partial_{\sigma}\,,
\end{equation}
appearing in the free-field equations that is transformed to momentum space. Equation (\ref{eq:differential-operator}) is
given in Feynman gauge. In principle the propagator can be obtained for a general three-momentum, i.e., for a generic
propagation direction. However the mathematical expressions are again quite involved and no underlying structure has
been found to express them in a compact way. Therefore the result is restricted to the special three-momentum
$\mathbf{k}=(0,0,k)$, which was the particular case considered in \secref{sec:limit-dispersion-relation} (but now
generalized to $k\in \mathbb{R}$). We obtain the following nonzero propagator coefficients:
\begin{subequations}
\label{eq:photon-propagator}
\begin{align}
\widehat{K}&=\frac{1}{k_0^4-2k_0^2k^2[1+2(\kappa^{0123})^2]+k^4}\,, \\[2ex]
\widehat{a}&=k_0^2-k^2\,,\quad \widehat{b}=4(\kappa^{0123})^2k_0^2\,,\quad \widehat{i}=-2\kappa^{0123}k_0 k\,, \\[2ex]
\widehat{l}&=\frac{4(\kappa^{0123})^2k_0^2}{k_0^2-k^2}\,,\quad \widehat{m}=-\frac{4(\kappa^{0123})^2k_0^3}{k_0^2-k^2}\,.
\end{align}
\end{subequations}
The remaining coefficients vanish. Three remarks of this result are in order. First, for $\kappa^{0123}=0$
it reduces to the standard photon propagator where $\widehat{a}\widehat{K}=1/(k_0^2-k^2)$. Second, the poles
of the scalar propagator part correspond to the modified physical dispersion relations that are obtained from
\eqref{eq:field-equations-modified-maxwell-theory} by setting $k_1=k_2=0$ and $k_3=k$:
\begin{equation}
\label{eq:dispersion-relation-special-case}
\omega_{1,2}(k)=|k|\sqrt{1+(\kappa^{0123})^2}\pm k\kappa^{0123}\,.
\end{equation}
Third, the pole of the unphysical coefficients $\widehat{l}$ and $\widehat{m}$ corresponds to the unphysical
dispersion relation $\omega_{0,3}(k)=k$ of the scalar and the longitudinal mode. This pole does not occur in the
physical coefficients $\widehat{a}$, $\widehat{b}$, and $\widehat{i}$.

The polarization vector for a special mode follows as a solution of the free field equations
(\ref{eq:field-equations-modified-maxwell-theory}) in momentum space. To obtain such a solution, $k_0$ has to
be replaced by the respective mode frequency and a gauge has to be chosen. Working in Lorenz
gauge $\partial_{\mu}A^{\mu}=0$, one obtains the following polarization vectors for the two physical photon
modes:
\begin{equation}
\label{eq:photon-polarization-vectors}
\varepsilon^{(1)\,\mu}=\frac{1}{\sqrt{2N'}}\begin{pmatrix}
0 \\
1 \\
1 \\
0 \\
\end{pmatrix}\,,\quad \varepsilon^{(2)\,\mu}=\frac{1}{\sqrt{2N''}}\begin{pmatrix}
0 \\
1 \\
-1 \\
0 \\
\end{pmatrix}\,.
\end{equation}
The normalizations $N'$ and $N''$ have to be chosen such that the 00-component of the energy-momentum
tensor (see, e.g., Eq.~(36) in \cite{ColladayKostelecky1998}) corresponds to $\omega_1$ for the first mode
and to $\omega_2$ for the second mode. Details of this approach applied to the isotropic sector of modified
Maxwell theory can be found in \cite{Kaufhold:2007}. For the birefringent case considered the normalizations
are given by:
\begin{equation}
\label{eq:normalization-polarization-vector}
N'=\frac{|k|}{\omega_1}\sqrt{1+(\kappa^{0123})^2}\,,\quad N''=\frac{|k|}{\omega_2}\sqrt{1+(\kappa^{0123})^2}\,.
\end{equation}
For vanishing Lorentz violation they correspond to the standard results $N'=N''=1$. The polarization vectors
of \eqref{eq:photon-polarization-vectors} are purely spacelike and orthogonal to each other. They provide
the physical basis vectors for the Fourier decomposition of the vector potential in terms of creation and
annihilation operators.

The next step is to construct the polarization tensors of the theory. The purpose of doing this is to relate
them to the propagator in a later part of the paper. A polarization tensor is a two-rank tensor that results
from combining the polarization vectors, i.e., it is given by $\overline{\varepsilon}^{(1)\,\mu}\varepsilon^{(1)\,\nu}$
for the first polarization mode and by $\overline{\varepsilon}^{(2)\,\mu}\varepsilon^{(2)\,\nu}$ for the second
mode. The bar above the first polarization vector in these expressions indicates complex conjugation. This
does not play a role here, though, as the polarization vectors can be chosen to be purely real. The intention
is to write the polarization tensors in a covariant form such as the propagator. This means that we make
an \textit{ansatz} similar to \eqref{eq:propagator-parity-odd-coeff} producing the following results for the
first polarization tensor:
\begin{subequations}
\begin{align}
\Pi^{\mu\nu}|_{\lambda=1}&\equiv \overline{\varepsilon}^{(1)\,\mu}\varepsilon^{(1)\,\nu}=\frac{1}{2N'}\Big\{-\eta^{\mu\nu}+\widehat{b}_1\,\xi_0^{\mu}\xi_0^{\nu}+\widehat{i}_1(\xi_1^{\mu}\xi_2^{\nu}+\xi_2^{\mu}\xi_1^{\nu}) \notag \\
&\phantom{{}\equiv {}\overline{\varepsilon}^{(1)\,\mu}\varepsilon^{(1)\,\nu}=\frac{1}{2N'}\Big\{-\eta^{\mu\nu}}+\widehat{l}_1\,k^{\mu}k^{\nu}+\widehat{m}_1\,(k^{\mu}\xi_0^{\nu}+\xi_0^{\mu}k^{\nu})\Big\}\Big|_{k_0=\omega_1}\,, \\[2ex]
\widehat{b}_1&=1-\frac{\omega_1^2}{k^2}\,,\quad \widehat{i}_1=1\,,\quad \widehat{l}_1=-\frac{1}{k^2}\,,\quad \widehat{m}_1=\frac{\omega_1}{k^2}\,.
\end{align}
\end{subequations}
The second polarization tensor is given by:
\begin{subequations}
\begin{align}
\Pi^{\mu\nu}|_{\lambda=2}&\equiv \overline{\varepsilon}^{(2)\,\mu}\varepsilon^{(2)\,\nu}=\frac{1}{2N''}\Big\{-\eta^{\mu\nu}+\widehat{b}_2\,\xi_0^{\mu}\xi_0^{\nu}+\widehat{i}_2(\xi_1^{\mu}\xi_2^{\nu}+\xi_2^{\mu}\xi_1^{\nu}) \notag \\
&\phantom{{}\equiv{}\overline{\varepsilon}^{(2)\,\mu}\varepsilon^{(2)\,\nu}=\frac{1}{2N''}\Big\{-\eta^{\mu\nu}}+\widehat{l}_2\,k^{\mu}k^{\nu}+\widehat{m}_2\,(k^{\mu}\xi_0^{\nu}+\xi_0^{\mu}k^{\nu})\Big\}\Big|_{k_0=\omega_2}\,, \\[2ex]
\widehat{b}_2&=1-\frac{\omega_2^2}{k^2}\,,\quad \widehat{i}_2=-1\,,\quad \widehat{l}_2=-\frac{1}{k^2}\,,\quad \widehat{m}_2=\frac{\omega_2}{k^2}\,.
\end{align}
\end{subequations}
Here the normalizations $N'$ and $N''$ are those of \eqref{eq:normalization-polarization-vector}. Four comments are
in order. First, note that each polarization tensor can be written in a covariant form, which is not possible in
standard QED when Feynman gauge is used. This behavior bears resemblance to the parity-odd nonbirefringent case of modified
Maxwell theory that was studied in \cite{Schreck:2011ai}. Possibly it occurs for all theories with birefringent photon
dispersion laws where it does not matter whether birefringence appears at first or at higher orders Lorentz violation.
Second, the sum of both polarization tensors reduces to the standard result\footnote{when contracted with a gauge-invariant
expression, i.e., dropping all terms that depend on the momentum four-vector $k^{\mu}$} (see, e.g.,
\cite{PeskinSchroeder1995}) for $\kappa^{0123}=0$:
\begin{equation}
\lim_{\kappa^{0123}\mapsto 0} \sum_{\lambda=1,2} \Pi^{\mu\nu}|_{\lambda}\Big|^{\mathrm{truncated}}=-\eta_{\mu\nu}\,,
\end{equation}
where ``truncated'' means that all terms proportional to the momentum four-vector $k^{\mu}$ have been dropped.
Third, each polarization tensor does not have a standard counterpart for $\kappa^{0123}=0$. Observe that the
coefficients $\widehat{b}_{1/2}$ are equal to zero for a vanishing Lorentz-violating parameter. This does not hold
for $\widehat{l}_{1/2}$ and $\widehat{m}_{1/2}$, in fact, but these are unphysical and the related terms do not play
a role for physical (i.e. gauge-invariant) quantities. However $\widehat{l}_{1/2}$ neither go to zero for $\kappa^{0123}=0$
nor they disappear when contracted with a gauge-invariant quantity. This is not surprising, though, since there
is no covariant expression for each tensor in standard QED as mentioned previously. Furthermore these terms are not
assumed to lead to any problems. For the parity-odd nonbirefringent modified Maxwell theory similar terms remain in the
limit of zero Lorentz violation. However, they were shown to have no importance for physics \cite{Schreck:2011ai}
due to the validness of the Ward identity. Fourth, the functional expressions for both polarization tensors are
very similar in contrast to the parity-odd case. For the latter sector the first polarization tensor had a completely
different structure compared to the second.

As a final remark in this section, the polarization vectors of the unphysical modes of \eqref{eq:dispersion-relations-unphysical}
can be obtained from the field equations as well. They are chosen as follows:
\begin{equation}
\varepsilon^{(0)\,\mu}=\begin{pmatrix}
1 \\
0 \\
0 \\
0 \\
\end{pmatrix}\,,\quad \varepsilon^{(3)\,\mu}=\begin{pmatrix}
0 \\
0 \\
0 \\
1 \\
\end{pmatrix}\,.
\end{equation}
The polarization vector of the scalar mode $\omega_0$ is a purely timelike four-vektor whereas the polarization vector
of the longitudinal mode $\omega_3$ points along the propagation direction of the wave. These properties mirror the
characteristics of standard QED.

\section{Validity of unitarity}
\label{sec:discussion-of-unitarity}

Every well-behaved quantum field theory ought to have to property of unitarity. This guarantees that no
probability is lost in physical processes, e.g., the scattering of particles.
So far, a bunch of examples for
Lorentz-violating theories have gathered for which unitarity was proven to hold. On the one hand, MCS-theory
was shown to be unitary as long as the preferred spacetime direction is spacelike~\cite{AdamKlinkhamer2001}. The isotropic and
the parity-odd nonbirefringent sector of modified Maxwell theory were proven to be unitary at tree level within some
range of the Lorentz-violating coefficients \cite{Klinkhamer:2010zs,Schreck:2011ai}. Furthermore there are recent
results on the unitarity of special Pais--Uhlenbeck models \cite{Lopez-Sarrion:2013kxa,Reyes:2013nca,Reyes:2013oca}.

On the other hand, it was shown that unitarity breaks down for timelike MCS-theory \cite{AdamKlinkhamer2001}.
These examples shall demonstrate that unitarity is likely to still be valid in a Lorentz-violating quantum field
theory. Nevertheless it is not something that should be taken for granted and, therefore, it will be investigated
in what follows.

There are two methods that can be used to analyze unitarity. The first is testing the property of reflection
positivity (see, e.g., \cite{Montvay:1994} for an introduction of this concept within the framework of lattice gauge
theory) and the second is proving the validness of the optical theorem. Reflection positivity is a condition that
has to hold for the two-point function of a quantum field theory in Euclidian space to ensure unitarity at
tree-level. It was used in \cite{AdamKlinkhamer2001,Klinkhamer:2010zs} to prove unitarity for certain regimes of
the theories considered in these papers.

The optical theorem relates the imaginary part of a forward scattering amplitude to the total cross section of
the particle physics process that results from performing all possible cuts of the respective amplitude. It is a
consequence of unitarity of the $S$-matrix of the underlying quantum field theory. We write the latter as
$S=\mathbf{1}+\mathrm{i}T$ with the identity $\mathbf{1}$ and the transfer matrix $T$ describing the scattering
of particles. If $S$ is not unitary it holds that
\begin{equation}
\mathbf{1}\neq SS^{\dagger}=\mathbf{1}+\mathrm{i}(T-T^{\dagger})+TT^{\dagger}\,,
\end{equation}
which results in $2\mathrm{Im}(T)\neq TT^{\dagger}$. For this reason a violation of unitarity is supposed to
be revealed in the optical theorem. Hence it can serve as a tool to show unitarity at each order of perturbation
theory. Furthermore the optical theorem can be used as a cross check for the polarization vectors and the photon
propagator what will become clear below.

To analyze unitarity, the optical theorem will be used in this paper where the considerations are restricted to tree-level.
As an ingredient a suitable particle physics process is needed. In accordance with \cite{Schreck:2011ai} the choice is the
annihilation of a left-handed electron $\mathrm{e}_L^-$ and a right-handed positron $\mathrm{e}_R^+$ to a modified
photon $\widetilde{\upgamma}$. Massless fermions are considered so that their helicity is physically well-defined.
We choose this process for a number of reasons. First, it is a relatively simple tree-level process including
a modified photon propagator. Second, it has no threshold, i.e., its kinematics are not that complicated. Third,
it involves fields that are not parity-invariant. This makes sense to reveal possible issues that may occur in
the context of parity violation as the particular birefringent theory considered violates parity. We neglect
the axial anomaly, which is of higher order with respect to the electromagnetic coupling constant.

\begin{figure}[t]
\centering
\begin{equation*}\label{eq:opt-theorem}
\hspace*{-5mm}
2\,\mathrm{Im}\left(\begin{array}{c}
\includegraphics{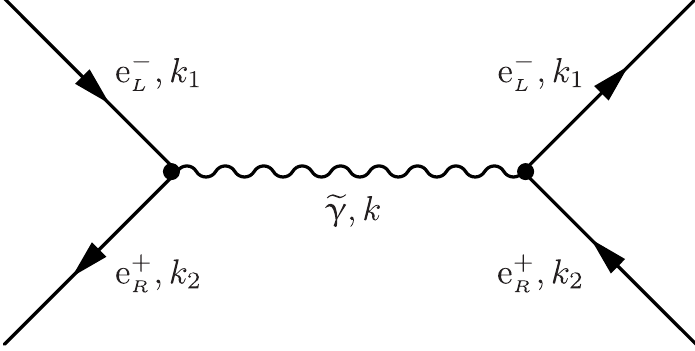}
\end{array}
\right)
\stackrel{?}{=}
\int \mathrm{d}\Pi_1 \left|\begin{array}{c}
\includegraphics{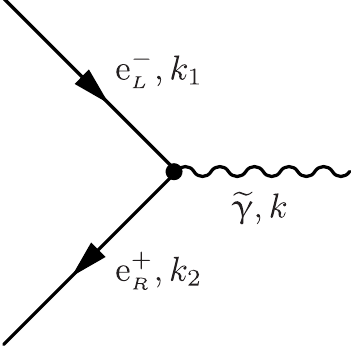}
\end{array}\right|^2\,.
\end{equation*}
\caption{Optical theorem tested for the process $\mathrm{e}^{-}_{L}\mathrm{e}^{+}_{R}\rightarrow \mathrm{e}^{-}_{L}\mathrm{e}^{+}_{R}$
within the birefringent QED extension considered. The respective kinematic variables are stated next to the particle symbols. The
infinitesimal one-particle phase space element for the process on the right-hand side of the equation is denoted as $\mathrm{d}\Pi_1$.}
\label{fig:optical-theorem}
\end{figure}

If the optical theorem is valid, the imaginary part of the forward scattering amplitude
$\mathcal{M}\equiv \mathcal{M}(\mathrm{e}^{-}_{L}\mathrm{e}^{+}_{R}\rightarrow \mathrm{e}^{-}_{L}\mathrm{e}^{+}_{R})$ is
related to the production cross-section of a modified photon from a left-handed electron and a right-handed positron.
The matrix element of the latter process will be denoted as
$\widehat{\mathcal{M}}\equiv \mathcal{M}(\mathrm{e}_L^-\mathrm{e}_R^+\rightarrow\widetilde{\upgamma})$.
The forward scattering amplitude reads as follows:
\begin{align}
\label{eq:forward-scattering-amplitude-optical-theorem}
\mathcal{M}&=\int \frac{\mathrm{d}^4k}{(2\pi)^4}\,\delta^{(4)}(k_1+k_2-k)\,
e^2\;\overline{u}(k_1)\gamma^{\lambda}\frac{\mathds{1}-\gamma_5}{2}v(k_2)\;
\overline{v}(k_2)\gamma^{\nu}\frac{\mathds{1}-\gamma_5}{2}u(k_1) \notag \displaybreak[0]\\
&\phantom{{}={}\int \frac{\mathrm{d}^4k}{(2\pi)^4}}\,\times\frac{1}{\widehat{K}^{-1}+\mathrm{i}\epsilon}\;
\big[+\widehat{a}\,\eta_{\nu\lambda}
+\widehat{b}\,\xi_{0,\nu}\xi_{0,\lambda}
+\widehat{i}\,(\xi_{1,\nu}\xi_{2,\lambda}+\xi_{2,\nu}\xi_{1,\lambda}) \notag \\
&\phantom{{}={}\int \frac{\mathrm{d}^4k}{(2\pi)^4}\,\times\frac{1}{\widehat{K}^{-1}+\mathrm{i}\epsilon}\;\big(}
+\widehat{l}\,k_{\nu}k_{\lambda}
+\widehat{m}\,(k_{\nu}\xi_{0,\lambda}+\xi_{0,\nu}k_{\lambda})
\big]\,,
\end{align}
where $e$ is the elementary charge, $u$, $v$, $\overline{u}$, and $\overline{v}$ are the respective Dirac
spinors, $\gamma_5=\mathrm{i}\gamma^0\gamma^1\gamma^2\gamma^3$ with the standard Dirac matrices $\gamma^{\mu}$
(for $\mu\in \{0,1,2,3\}$), and $\mathds{1}$ is the unit matrix in spinor space. The kinematical variables are
shown in \figref{fig:optical-theorem}. The four-dimensional $\delta$-function ensures total four-momentum conservation.
The photon propagator with the respective propagator coefficients is taken from \eqref{eq:photon-propagator}. Note
that for a proper treatment of the propagator poles in Minkowski spacetime the usual $\mathrm{i}\epsilon$-procedure
is applied for the physical poles. The scalar part $\widehat{K}$ of the propagator is written in the following form:
\begin{align}
\frac{1}{\widehat{K}^{-1}+\mathrm{i}\epsilon}&=\frac{1}{k_0^4-2k_0^2k^2[1+2(\kappa^{0123})^2]+k^4+\mathrm{i}\epsilon} \notag \\
&=\frac{1}{(k^0-\omega^++\mathrm{i}\epsilon)(k^0+\omega^+-\mathrm{i}\epsilon)(k^0-\omega^-+\mathrm{i}\epsilon)(k^0+\omega^--\mathrm{i}\epsilon)}\,,
\end{align}
where the factorization is done with respect to the propagator poles. Terms of quadratic and higher order in
the infinitesimal parameter $\epsilon$ are dropped. Contrary to the standard photon propagator there appear four poles,
where the positive ones are given by:
\begin{equation}
\omega^+=\sqrt{1+(\kappa^{0123})^2}|k|+\kappa^{0123}k\,,\quad \omega^-=\sqrt{1+(\kappa^{0123})^2}|k|-\kappa^{0123}k\,.
\end{equation}
This is characteristic for a birefringent quantum field theory and it is also the case for the parity-odd
nonbirefringent sector \cite{Schreck:2011ai} since in the latter birefringence still occurs at quadratic
order in the Lorentz-violating coefficients. However in the parity-odd case the propagator structure was
more involved than for the birefringent case considered here. Now, $\omega^+$,
$\omega^-$ are the physical poles and $-\omega^+$, $-\omega^-$ are their negative counterparts. From the
mathematical point of view propagators are distributions. Due to the $\mathrm{i}\epsilon$-prescription the
following relations then hold for the physical poles:
\begin{subequations}
\label{eq:propagator-relations-principal-value}
\begin{align}
\label{eq:propagator-relations-principal-value-1}
\frac{1}{k^0-\omega^++\mathrm{i}\epsilon}&=\mathcal{P}\frac{1}{k^0-\omega^+}-\mathrm{i}\pi\delta(k^0-\omega^+)\,, \\[2ex]
\label{eq:propagator-relations-principal-value-2}
\frac{1}{k^0-\omega^-+\mathrm{i}\epsilon}&=\mathcal{P}\frac{1}{k^0-\omega^-}-\mathrm{i}\pi\delta(k^0-\omega^-)\,,
\end{align}
\end{subequations}
where $\mathcal{P}$ denotes the principal value. The first part of Eqs.~(\ref{eq:propagator-relations-principal-value-1})
and (\ref{eq:propagator-relations-principal-value-2}), respectively,
is purely real. The second part is imaginary and due to the $\delta$-function it forces the zeroth four-momentum
component to be equal to the respective physical photon frequencies. The negative counterparts of the poles will
not play any role in the calculation below because of total four-momentum conservation.

Using these results, the $k^0$-integration in \eqref{eq:forward-scattering-amplitude-optical-theorem} can
be performed. According to \eqref{eq:propagator-relations-principal-value} each physical pole delivers a
contribution. If we are only interested in the imaginary part, we can drop terms involving the principal
values and replace $k^0$ by the photon frequency for each term. In doing so, the following intermediate results
are useful:
\begin{subequations}
\begin{align}
\frac{(\omega^+)^2-k^2}{2\omega^+[(\omega^+)^2-(\omega^-)^2]}&=\frac{1}{4|k|\sqrt{1+(\kappa^{0123})^2}}=\frac{1}{4\omega^+N'}\,, \\[2ex]
\frac{(\omega^-)^2-k^2}{2\omega^-[(\omega^-)^2-(\omega^+)^2]}&=\frac{1}{4|k|\sqrt{1+(\kappa^{0123})^2}}=\frac{1}{4\omega^-N''}\,.
\end{align}
\end{subequations}
Finally we obtain for the first photon mode
\begin{subequations}
\begin{align}\label{eq:optical-theorem-mode-1}
2\,\mathrm{Im}(\mathcal{M})\big|_{\lambda=1}&=\int
\frac{\mathrm{d}^3k}{(2\pi)^{3}\,2\omega^+}\,\delta^{(4)}(k_1+k_2-k) \notag \\
&\hspace{1.5cm}\times e^2\;\overline{u}(k_1)\gamma^{\nu}\frac{\mathds{1}-\gamma_5}{2}v(k_2)\;
\overline{v}(k_2)\gamma^{\mu}\frac{\mathds{1}
-\gamma_5}{2}u(k_1) \notag \\
&\hspace{1.5cm}\times
\frac{1}{2N'}\;\Big\{-\eta_{\mu\nu}+\widehat{b}_1\,\xi_0^{\mu}\xi_0^{\nu}
+\widehat{i}_1\,(\xi_1^{\mu}\xi_2^{\nu}+\xi_2^{\mu}\xi_1^{\nu})\Big\}
\notag \displaybreak[0]\\
&=\int \frac{\mathrm{d}^3k}{(2\pi)^{3}\,2\omega^+}\,\delta^{(4)}(k_1+k_2-k)\,
(\widehat{\mathcal{M}}^{\,\dagger})^{\nu}(\widehat{\mathcal{M}})^{\mu}\Big(\Pi_{\mu\nu}|_{\lambda=1}\Big)
\notag \displaybreak[0]\\[1mm]
&=\int
\frac{\mathrm{d}^3k}{(2\pi)^{3}\,2\omega^+}\,\delta^{(4)}(k_1+k_2-k)\left.
|\widehat{\mathcal{M}}|^2\right|_{\lambda=1}\,,
\end{align}
with
\begin{equation}
\left.\widehat{\mathcal{M}}\right|_{\lambda=1}\equiv \varepsilon^{(1)}_{\mu}(k)(\widehat{\mathcal{M}})^{\mu}(k)\,.
\end{equation}
\end{subequations}
Terms that involve at least one four-momentum in the tensor structure of the propagator can be dropped, if the Ward
identity is taken into account. The calculation for the second mode is completely analogous:
\begin{subequations}
\begin{align}\label{eq:optical-theorem-mode-2}
2\,\mathrm{Im}(\mathcal{M})\big|_{\lambda=2}&=\int
\frac{\mathrm{d}^3k}{(2\pi)^{3}\,2\omega^-}\,\delta^{(4)}(k_1+k_2-k) \notag \\
&\hspace{1.5cm}\times e^2\;\overline{u}(k_1)\gamma^{\nu}\frac{\mathds{1}-\gamma_5}{2}v(k_2)\;
\overline{v}(k_2)\gamma^{\mu}\frac{\mathds{1}
-\gamma_5}{2}u(k_1) \notag \\
&\hspace{1.5cm}\times
\frac{1}{2N''}\;\Big\{-\eta_{\mu\nu}+\widehat{b}_2\,\xi_0^{\mu}\xi_0^{\nu}
+\widehat{i}_2\,(\xi_1^{\mu}\xi_2^{\nu}+\xi_2^{\mu}\xi_1^{\nu})\Big\}
\notag \displaybreak[0]\\
&=\int
\frac{\mathrm{d}^3k}{(2\pi)^{3}\,2\omega^-}\,\delta^{(4)}(k_1+k_2-k)\left.
|\widehat{\mathcal{M}}|^2\right|_{\lambda=2}\,,
\end{align}
where
\begin{equation}
\left.\widehat{\mathcal{M}}\right|_{\lambda=2}\equiv \varepsilon^{(2)}_{\mu}(k)(\widehat{\mathcal{M}})^{\mu}(k)\,.
\end{equation}
\end{subequations}
To summarize, the validity of the optical theorem for the birefringent modified Maxwell theory considered was shown at tree-level
for a particular scattering process. This process was very generic and the formal proof did not rely on equations and kinematical
relations that are specific to this case only. The input was the modified photon propagator, the photon polarization vectors,
the treatment of the propagator poles via the usual $\mathrm{i}\epsilon$-prescription, and the Ward identity. Therefore we conjecture
that the modified QED is unitary at tree-level --- at least for the choice $\mathbf{k}=(0,0,k)$ of the three-momentum. An
analytical proof of unitarity for a general four-momentum is difficult due to the complicated structure of the dispersion law, the
polarization vectors, and the propagator and it is an interesting open problem for future studies. However, in
App.~\ref{eq:numerical-check-optical-theorem} a numerical check of the optical theorem is presented for a spatial momentum with
nonvanishing components. In this context the propagator and the polarization tensors are obtained for a general spatial momentum
$\mathbf{k}=(k_1,k_2,k_3)$. A curiosity is that each power of $\kappa^{0123}$, which appears in the tensor coefficients of
Eqs.~(\ref{eq:general-polarization-tensor-coefficient-1}) -- (\ref{eq:general-polarization-tensor-coefficient-10}), the normalization of \eqref{eq:general-polarization-vector-normalization}, and the propagator coefficients given by Eqs.~(\ref{eq:general-propagator-coefficient-1}) -- (\ref{eq:general-propagator-coefficient-10}), is multiplied with a number that is a power of two. This may indicate that the modified
theory is based on a mathematical structure, which is not understood at present.

\section{Microcausality}
\label{sec:discussion-of-microcausality}

In \secref{sec:dispersion-relations-classical-theory} we analyzed the phase and group velocities of the physical photon modes. The
upshot was that within the classical theory at least one mode always propagates with a superluminal velocity. For this reason
we are interested in studying causality at the quantum level. This is done by evaluating the commutator of two vector potentials at
distinct spacetime points $x$ and $y$. Such a procedure helps to understand how $x$ and $y$ are causally related to each other. If the
commutator vanishes the points are causally disconnected. This means that a quantum mechanical measurement of physical fields at $x$
does not have any influence on the measurement of an observable at $y$. In other words, between the two points, information cannot
be exchanged via a (modified) light signal. If the commutator does not vanish, $x$ and $y$ are causally connected, i.e., information can
be transported from one to the other point. By evaluating this commutator, spacetime regions can be determined that are causally
connected, which will give us insight on the issue of microcausality. For standard QED such calculations were performed by Pauli
and Jordan \cite{JordanPauli1928,Heitler1954}.

Due to translational invariance the commutator can be written such that it only depends on the spacetime coordinates of a single
point $x$:
\begin{equation}
[A^{\mu}(y),A^{\nu}(z)]=[A^{\mu}(y-z),A^{\nu}(0)]\equiv [A^{\mu}(x),A^{\nu}(0)]=\mathrm{i}\theta^{\mu\nu}D(x)\,.
\end{equation}
Its result is made up of a two-rank tensor $\theta^{\mu\nu}$, respecting the tensor structure of the commutator, and a scalar
commutator function $D(x)$. The causal structure of the theory is determined by the latter scalar function that will be computed
in what follows. To make the calculation feasible the photon modes with the dispersion relations
\begin{equation}
\omega^+=\sqrt{1+(\kappa^{0123})^2}|k|+\kappa^{0123}k\,,\quad \omega^-=\sqrt{1+(\kappa^{0123})^2}|k|-\kappa^{0123}k\,,
\end{equation}
which were also considered in the previous section, will be examined. The scalar commutator function is the Fourier transform of
the scalar propagator function \cite{AdamKlinkhamer2001}. Note that contrary to the latter reference we will compute the
Fourier transform of the product $\widehat{a}\widehat{K}$ of \eqref{eq:photon-propagator} and not only of $\widehat{K}$, which
will allow a comparison to the standard result. What matters for the causal properties of the theory is, indeed, the pole structure
of $\widehat{K}$, which
corresponds to the pole structure of $\widehat{a}\widehat{K}$. The integration runs over both $k^0$ and the spatial momentum
component. The integral over $k^0$ is performed in the complex $k^0$-$k$-plane where all poles are encircled along a closed
contour $C$ in counterclockwise direction. After that the resulting expression is integrated over the spatial momentum component
along the real axis. Hence we have to compute the following:
\begin{align}
\label{eq:starting-point-commutator-function}
D(x^0,x^3)&=\frac{1}{(2\pi)^2}\oint_C \mathrm{d}k_0 \int_{-\infty}^{\infty} \mathrm{d}k\,\frac{k_0^2-k^2}{k_0^4-2k_0^2k^2[1+2(\kappa^{0123})^2]+k^4}\exp(-\mathrm{i}kx) \notag \\
&=\frac{1}{(2\pi)^2}\oint_C \mathrm{d}k^0 \int_{-\infty}^{\infty} \mathrm{d}k\,\frac{\big[(k^0)^2-k^2\big]\exp(-\mathrm{i}k^0x^0+\mathrm{i}kx^3)}{(k^0-\omega^+)(k^0+\omega^+)(k^0-\omega^-)(k^0+\omega^-)}\,.
\end{align}
Note that the commutator function is defined in two rather than four dimensions in contrast to the investigations performed in
\cite{AdamKlinkhamer2001,Klinkhamer:2010zs,Schreck:2011ai}. Furthermore the sign convention in the complex exponential function
is chosen differently from the latter two references. The detailed steps of the calculation are relegated to
App.~\ref{sec:calculation-commutator-function}. Its result then reads as follows:
\begin{subequations}
\begin{align}
\label{eq:final-result-commutator-function}
&D(x^0,x^3)=\frac{1}{4\mathcal{A}}\mathrm{sgn}(x^0)\left\{\theta\left[(\mathcal{B}x^0)^2-(x^3)^2\right]+\theta\left[(\widetilde{\mathcal{B}}x^0)^2-(x^3)^2\right]\right\}\,, \\[2ex]
&\mathcal{B}=\mathcal{A}+\kappa^{0123}\,,\quad \widetilde{\mathcal{B}}=\mathcal{A}-\kappa^{0123}\,,\quad \mathcal{A}=\sqrt{1+(\kappa^{0123})^2}\,,
\end{align}
with the sign function and the Heaviside step function:
\begin{equation}
\label{eq:definition-heaviside-and-sign-function}
\mathrm{sgn}(x)=\left\{\begin{array}{rcc}
1 & \text{for} & x>0\,, \\
0 & \text{for} & x=0\,, \\
-1 & \text{for} & x<0\,, \\
\end{array}
\right.\quad \theta(x)=\left\{\begin{array}{lcl}
1 & \text{for} & x>0\,, \\
1/2 & \text{for} & x=0\,, \\
0 & \text{for} & x<0\,. \\
\end{array}
\right.
\end{equation}
\end{subequations}
The standard result for the commutator function of a scalar field with mass $m$ in two dimensions can be found, e.g., in
\cite{Heinzl:1999fz,Chakrabarti:1999cg}. For $\kappa^{0123}\mapsto 0$ \eqref{eq:final-result-commutator-function} corresponds to
the result given in the latter two references for $m=0$. The only difference is a global minus sign, which comes from the different
convention that is used here. The structure of the commutator function is completely different compared to the standard result in
four dimensions, which involves $\delta$-functions rather than $\theta$-functions. The arguments of the $\theta$-functions can be
interpreted as the modified nullcones of the two physical photon modes in configuration space:
\begin{subequations}
\label{eq:nullcones-birefringent-modmax}
\begin{align}
(x^0)^2-\frac{(x^3)^2}{\widetilde{c}_1^2}&=0\,,\quad \widetilde{c}_1=\mathcal{A}+\kappa^{0123}\,, \\[2ex]
(x^0)^2-\frac{(x^3)^2}{\widetilde{c}_2^2}&=0\,,\quad \widetilde{c}_2=\mathcal{A}-\kappa^{0123}\,,
\end{align}
\end{subequations}
where the constants $\widetilde{c}_1$ and $\widetilde{c}_2$ have the dimension of velocities. Note that the phase velocities of the
modes are given by:
\begin{equation}
\label{eq:phase-velocities-microcausality}
v_{\mathrm{ph},1}=\frac{\omega_1}{k}=\mathcal{A}\,\mathrm{sgn}(k)+\kappa^{0123}\,,\quad v_{\mathrm{ph},2}=\frac{\omega_2}{k}=\mathcal{A}\,\mathrm{sgn}(k)-\kappa^{0123}\,.
\end{equation}
Hence what appears as $\widetilde{c}_1$ and $\widetilde{c}_2$ in \eqref{eq:nullcones-birefringent-modmax} are the values of the phase
velocities given in \eqref{eq:phase-velocities-microcausality} for $k>0$. Due to birefringence there appear two modified nullcones,
which merge to the standard
one for $\kappa^{0123}=0$. The scalar commutator function of \eqref{eq:final-result-commutator-function} tells us that the commutator
vanishes outside the modified nullcones. For this reason, measurements of quantum mechanical observables at two spacetime
points can only influence each other if the respective points can be connected by a world line lying on or within one of the two
nullcones.\footnote{General modified nullcones can intersect each other (see the discussion at the end of \secref{sec:limit-dispersion-relation}
on the degeneracy of dispersion laws in momentum space). So in principle, points can as well be connected by a world line lying within
such an intersection of both nullcones.} Hence information can only
propagate along or inside these cones, which is the crucial property for a microcausal theory. The birefringent sector considered
is, therefore, microcausal, at least in the $x^0$-$x^3$-plane. The proof for the total configuration space, which would involve the
general dispersion relations of \eqref{eq:dispersion-relations-birefringent}, is a challenge and an interesting open problem.

\subsection{Effective metrics and bi-metric theories}

Some final comments shall be made on effective metrics that appear in this framework. In certain Lorentz-violating theories such metrics
can be constructed so that the modified photon four-momentum squared (or the nullcone coordinate vector squared) is zero with
respect to the suitable metric. For example, in \cite{Betschart:2008yi} effective metrics were introduced for certain nonbirefringent
cases of modified Maxwell theory that were coupled to a gravitational background.

First of all, for the birefringent case restricted to the particular three-momentum $\mathbf{k}=(0,0,k)$ we define effective metrics
$\widetilde{g}^{\mu\nu}_i$ ($i=1$, 2) in momentum space with $\widetilde{g}^{\mu\nu}_ik_{\mu}k_{\nu}=0$ for the two physical photon
modes. These metrics are given by:
\begin{equation}
(\widetilde{g}^{\mu\nu})=\left\{\begin{array}{lcl}
\mathrm{diag}\left(1,-1,-1,-\left[\mathcal{A}+\kappa^{0123}\right]^2\right)\equiv (\widetilde{g}^{\mu\nu}_1) & \text{for} & k^0=\omega_1,\,k\geq 0 \text{ or } \\
 & & k^0=\omega_2,\,k<0\,, \\
\mathrm{diag}\left(1,-1,-1,-\left[\mathcal{A}-\kappa^{0123}\right]^2\right)\equiv (\widetilde{g}^{\mu\nu}_2) & \text{for} & k^0=\omega_1,\,k< 0 \text{ or } \\
 & & k^0=\omega_2,\,k\geq 0\,. \\
\end{array}
\right.
\end{equation}
Note that the index $i$ of the effective metric does not necessarily correspond to the index $\lambda=1$, 2 of the physical mode. Moreover,
analogue effective metrics $g_{\mu\nu}^{(i)}$ can be introduced in configuration space such that $g_{\mu\nu}^{(i)}x^{\mu}x^{\nu}=0$
for both physical modes. These are often called nullcone metrics. According to the final discussion of the previous section,
for $x_3\geq 0$ the following results are obtained:
\begin{subequations}
\begin{equation}
\label{eq:nullcone-metric-1}
(g_{\mu\nu}^{(1)})=\mathrm{diag}\left(1,-1,-1,-\frac{1}{\left[\mathcal{A}+\kappa^{0123}\right]^2}\right)\,,
\end{equation}
for the first modified nullcone and
\begin{equation}
\label{eq:nullcone-metric-2}
(g_{\mu\nu}^{(2)})=\mathrm{diag}\left(1,-1,-1,-\frac{1}{\left[\mathcal{A}-\kappa^{0123}\right]^2}\right)\,,
\end{equation}
\end{subequations}
for the second nullcone. It is evident that each nullcone metric is the inverse of the corresponding effective metric in
momentum space. Hence in principle the theory considered in this paper is a bi-metric quantum field theory in a flat spacetime.
Theories characterized by even multiple nullcone metrics were studied in, e.g., \cite{Drummond:2013ida}. For a bi-metric theory
with the metrics $g_{\mu\nu}^{(1)}$ and $g_{\mu\nu}^{(2)}$ an additional metric $\widehat{g}_{\mu\nu}$ can be introduced
interpolating between them:
\begin{equation}
\widehat{g}_{\mu\nu}\equiv (1-u)g_{\mu\nu}^{(1)}+ug_{\mu\nu}^{(2)}\,,\quad 0\leq u\leq 1\,.
\end{equation}
Under certain circumstances this interpolating metric becomes singular indicating problems with causality. However such a behavior
does not occur as long as there are vectors that are timelike with respect to the two metrics $g_{\mu\nu}^{(i)}$, which is the case
when both nullcones overlap (see figure 1(i) and (ii) in the latter reference). This criterium is fulfilled here as the second nullcone
characterized by the metric (\ref{eq:nullcone-metric-2}) lies inside of the first nullcone with the metric (\ref{eq:nullcone-metric-1})
for $\kappa^{0123}>0$ and vice versa for $\kappa^{0123}<0$.

\section{Conclusions and outlook}

In this article birefringent modified Maxwell theory coupled to a Dirac theory of standard spin-1/2 particles was examined. Thereby a
special emphasis was put on the respective quantum field theoretic properties. To keep calculations feasible the birefringent photon
sector was restricted to one nonzero Lorentz-violating coefficient. Within this setup, the modified photon dispersion relations, the
propagator, and the polarization vectors were obtained. For both the propagator and the polarization vectors the three-momentum was
chosen to point along the third axis of the coordinate system to simplify the complicated functional structure of the expressions.

With these results both unitarity (at tree-level) and microcausality of the respective birefringent QED were demonstrated to be valid.
Besides, an issue of the theory for a negative Lorentz-violating coefficient was notified that also occurs for certain Finsler
structures.

To the best knowledge of the author this is one of the first attempts to understand this special birefringent modified QED. The articles
\cite{Colladay:2013dra,Colladay:2014dua}, which were published recently, address certain peculiarities of Gupta-Bleuler quantization in the
context of birefringent modified Maxwell theory. It was shown that a nonvanishing photon mass has to be introduced such that quantization
is still possible. As examples in the latter references, frameworks with other nonvanishing Lorentz-violating coefficients were examined
in comparison to the one considered in the current article. It is also worth mentioning that in \cite{Kruglov:2007zr} a birefringent theory
is considered that is not only quadratic but cubic in the electromagnetic field strength tensor.

The current article can provide a basis for future studies of quantum field theories that are based on a birefringent photon sector.
This is not necessarily restricted to an analysis of modified Maxwell theory. However, especially for modified Maxwell theory there
are still quite some problems that may be investigated in future. For example, understanding the full mathematical structure of the
theory considered may help to write the general propagator and the polarization vectors in a compact form. Then at least the proof
of unitarity at tree-level with the help of the optical theorem may be possible for a general three-momentum. Furthermore, there are
nine remaining birefringent Lorentz-violating coefficients whose corresponding theories await investigation.

\section{Acknowledgments}

It is a pleasure to thank V.~A.~Kosteleck\'{y} for reading the manuscript and making very useful suggestions. Furthermore the author
is indebted to the anonymous referee for interesting comments. This work was performed with financial support from the
\textit{Deutsche Akademie der Naturforscher Leopoldina} within Grant No. LPDS 2012-17.

\begin{appendix}
\numberwithin{equation}{section}

\section{Calculation of the commutator function}
\label{sec:calculation-commutator-function}

In this appendix the integral in \eqref{eq:starting-point-commutator-function} will be evaluated in detail. Starting with the latter
equation, the residue theorem leads to:
\begin{align}
\label{eq:commutator-function-after-residue-theorem}
D(x^0,x^3)&=\frac{\mathrm{i}}{2\pi} \int_{-\infty}^{\infty} \mathrm{d}k\,\exp(\mathrm{i}kx^3)\left\{\frac{(\omega^+)^2-k^2}{2\omega^+[(\omega^+)^2-(\omega^-)^2]}\exp(-\mathrm{i}\omega^+x^0)\right. \notag \\
&\phantom{{}={}\frac{\mathrm{i}}{2\pi} \int_{-\infty}^{\infty} \mathrm{d}k\,\exp(\mathrm{i}kx^3)\Big\{}+\frac{(\omega^+)^2-k^2}{-2\omega^+[(\omega^+)^2-(\omega^-)^2]}\exp(\mathrm{i}\omega^+x^0) \notag \\
&\phantom{{}={}\frac{\mathrm{i}}{2\pi} \int_{-\infty}^{\infty} \mathrm{d}k\,\exp(\mathrm{i}kx^3)\Big\{}+\frac{(\omega^-)^2-k^2}{2\omega^-[(\omega^-)^2-(\omega^+)^2]}\exp(-\mathrm{i}\omega^-x^0) \notag \\
&\phantom{{}={}\frac{\mathrm{i}}{2\pi} \int_{-\infty}^{\infty} \mathrm{d}k\,\exp(\mathrm{i}kx^3)\Big\{}\left.+\frac{(\omega^-)^2-k^2}{-2\omega^-[(\omega^-)^2-(\omega^+)^2]}\exp(\mathrm{i}\omega^-x^0)\right\}\,.
\end{align}
In two (rather than four) dimensions it is more convenient to consider the separate complex exponential functions instead of combining
them to real trigonometric functions. This will become clearer below. The expressions in the curly brackets can be further evaluated to
give:
\begin{subequations}
\label{eq:intermediate-terms}
\begin{align}
\frac{(\omega^-)^2-k^2}{\omega^-[(\omega^+)^2-(\omega^-)^2]}&=-\frac{1}{2|k|\sqrt{1+(\kappa^{0123})^2}}\,, \\[2ex] \frac{(\omega^+)^2-k^2}{\omega^+[(\omega^+)^2-(\omega^-)^2]}&=\frac{1}{2|k|\sqrt{1+(\kappa^{0123})^2}}\,.
\end{align}
\end{subequations}
With \eqref{eq:intermediate-terms} and the definition $\mathcal{A}\equiv \sqrt{1+(\kappa^{0123})^2}$ the result of
\eqref{eq:commutator-function-after-residue-theorem} can be simplified:
\begin{align}
D(x^0,x^3)&=\frac{\mathrm{i}}{2\pi}\frac{1}{4\mathcal{A}}\int_{-\infty}^{\infty} \mathrm{d}k\,\frac{1}{|k|}\exp(\mathrm{i}kx^3)\left[\exp(-\mathrm{i}\omega^+x^0)-\exp(\mathrm{i}\omega^+x^0)\right. \notag \\
&\phantom{{}={}\frac{\mathrm{i}}{2\pi}\frac{1}{4\mathcal{A}}\int_{-\infty}^{\infty} \mathrm{d}k\,\frac{1}{|k|}\exp(\mathrm{i}kx^3)\big[}\left.+\exp(-\mathrm{i}\omega^-x^0)-\exp(\mathrm{i}\omega^-x^0)\right]\,.
\end{align}
To avoid absolute values of $k$ in the integrand, the integration domain is divided into the region of positive $k$ and negative $k$.
For $k>0$ one obtains:
\begin{subequations}
\begin{align}
\omega^+&=\mathcal{A}k+k\kappa^{0123}=(\mathcal{A}+\kappa^{0123})k\equiv \mathcal{B}k\,, \\[2ex]
\omega^-&=\mathcal{A}k-k\kappa^{0123}=(\mathcal{A}-\kappa^{0123})k\equiv \widetilde{\mathcal{B}}k\,.
\end{align}
\end{subequations}
The first part of the commutator function associated with positive $k$ is denoted as $D_1$:
\begin{align}
\label{eq:commutator-function-1}
D_1(x^0,x^3)&=\frac{\mathrm{i}}{2\pi}\frac{1}{4\mathcal{A}}\int_0^{\infty} \mathrm{d}k\,\frac{1}{k}\left\{\exp\left[\mathrm{i}k(x^3-\mathcal{B}x^0)\right]-\exp\left[\mathrm{i}k(x^3+\mathcal{B}x^0)\right]\right. \notag \\
&\phantom{{}={}\frac{\mathrm{i}}{2\pi}\frac{1}{4\mathcal{A}}\int_0^{\infty} \mathrm{d}k\,\frac{1}{k}\big\{}\left.+\exp\left[\mathrm{i}k(x^3-\widetilde{\mathcal{B}}x^0)\right]-\exp\left[\mathrm{i}k(x^3+\widetilde{\mathcal{B}}x^0)\right]\right\}\,.
\end{align}
For $k<0$ the modified dispersion laws can be written as follows:
\begin{subequations}
\begin{align}
\omega^+&=-\mathcal{A}k+k\kappa^{0123}=-(\mathcal{A}-\kappa^{0123})k=-\widetilde{\mathcal{B}}k\,, \\[2ex]
\omega^-&=-\mathcal{A}k-k\kappa^{0123}=-(\mathcal{A}+\kappa^{0123})k=-\mathcal{B}k\,.
\end{align}
\end{subequations}
The second part of the scalar commutator function associated with negative values of $k$ is denoted as $D_2$ and reads:
\begin{align}
\label{eq:commutator-function-2}
D_2(x^0,x^3)&=\frac{\mathrm{i}}{2\pi}\frac{1}{4\mathcal{A}}\int_{-\infty}^0 \mathrm{d}k\,\frac{1}{-k}\left\{\exp\left[\mathrm{i}k(x^3+\widetilde{\mathcal{B}}x^0)\right]-\exp\left[\mathrm{i}k(x^3-\widetilde{\mathcal{B}}x^0)\right]\right. \notag \\
&\phantom{{}={}\frac{\mathrm{i}}{2\pi}\frac{1}{4\mathcal{A}}\int_{-\infty}^0 \mathrm{d}k\,\frac{1}{-k}\Big\{}\left.+\exp\left[\mathrm{i}k(x^3+\mathcal{B}x^0)\right]-\exp\left[\mathrm{i}k(x^3-\mathcal{B}x^0)\right]\right\}\,.
\end{align}
The total commutator function follows by adding \eqref{eq:commutator-function-1} and \eqref{eq:commutator-function-2}. To avoid the pole
at $k=0$ the replacement $k\mapsto k+\mathrm{i}\epsilon$ with $\epsilon=0^+$ is performed. This sign of $\epsilon$ is chosen according
to the usual prescription that is applied to define the Feynman propagator. One then obtains:
\begin{align}
D(x^0,x^3)&=D_1(x^0,x^3)+D_2(x^0,x^3) \notag \\
&=-\frac{1}{2\pi\mathrm{i}}\frac{1}{4\mathcal{A}}\int_{-\infty}^{\infty} \mathrm{d}k\,\frac{1}{k+\mathrm{i}\epsilon}\left\{\exp\left[\mathrm{i}k(x^3-\mathcal{B}x^0)\right]-\exp\left[\mathrm{i}k(x^3+\mathcal{B}x^0)\right]\right. \notag \\
&\phantom{{}={}-\frac{1}{2\pi\mathrm{i}}\frac{1}{4\mathcal{A}}\int_{-\infty}^{\infty} \mathrm{d}k\,\frac{1}{k+\mathrm{i}\epsilon}\big\{}\left.+\exp\left[\mathrm{i}k(x^3-\widetilde{\mathcal{B}}x^0)\right]-\exp\left[\mathrm{i}k(x^3+\widetilde{\mathcal{B}}x^0)\right]\right\}\,.
\end{align}
Since the standard commutator function is a distribution, $D(x^0,x^3)$ will be interpreted as a distribution as well that is supposed
to act on a smooth function $f(x^0,x^3)$. Each of the four terms above can be written as a Heaviside step function $\theta(x)$, which
is defined in \eqref{eq:definition-heaviside-and-sign-function}.
The following integral representations of $\theta(x)$ turn out to be useful in the current context:
\begin{equation}
\lim_{\epsilon\mapsto 0^+} \frac{1}{2\pi\mathrm{i}}\int_{-\infty}^{\infty} \mathrm{d}k\,\frac{\exp(\mathrm{i}kx)}{k-\mathrm{i}\epsilon}=\theta(x)\,,\quad \lim_{\epsilon\mapsto 0^+} \frac{1}{2\pi\mathrm{i}}\int_{-\infty}^{\infty} \mathrm{d}k\,\frac{\exp(\mathrm{i}kx)}{k+\mathrm{i}\epsilon}=-\theta(-x)\,.
\end{equation}
This leads to the final result for the commutator function:
\begin{align}
D(x^0,x^3)&=-\frac{1}{4\mathcal{A}}\left[-\theta(\mathcal{B}x^0-x^3)+\theta(-\mathcal{B}x^0-x^3)-\theta(\widetilde{\mathcal{B}}x^0-x^3)+\theta(-\widetilde{\mathcal{B}}x^0-x^3)\right] \notag \\
&=\frac{1}{4\mathcal{A}}\mathrm{s
gn}(x^0)\left\{\theta\left[(\mathcal{B}x^0)^2-(x^3)^2\right]+\theta\left[(\widetilde{\mathcal{B}}x^0)^2-(x^3)^2\right]\right\} \notag \\
&=\frac{1}{8\mathcal{A}}\left[\mathrm{sgn}(\mathcal{B}x^0+x^3)+\mathrm{sgn}(\mathcal{B}x^0-x^3)\right. \notag \\
&\phantom{{}={}\frac{1}{8\mathcal{A}}\big[}\left.+\,\mathrm{sgn}(\widetilde{\mathcal{B}}x^0+x^3)+\mathrm{sgn}(\widetilde{\mathcal{B}}x^0-x^3)\right]\,,
\end{align}
with the sign function defined in \eqref{eq:definition-heaviside-and-sign-function}.

\section{Numerical check of the optical theorem}
\label{eq:numerical-check-optical-theorem}

In \secref{sec:discussion-of-unitarity} the validity of the optical theorem was shown for a particular scattering process. This was done
for the momentum configuration $\mathbf{k}=(0,0,k)$ due to the complexities of the modified photon polarization vectors, the propagator,
and the dispersion laws. In the current section the respective results are given for a general spatial momentum $\mathbf{k}=(k_1,k_2,k_3)$.
Note that in contrast to Secs.~\ref{sec:propagator-polarization-vectors} -- \ref{sec:discussion-of-microcausality} plus
App.~\ref{sec:calculation-commutator-function}, $k^2=k^{\mu}k_{\mu}$ in the current section. First of all, the polarization vector
components in Lorenz gauge are chosen in the following form:
\begin{align}
\varepsilon^0&=\frac{2\kappa^{0123}\left(k^2k_2-4\kappa^{0123}\omega k_1k_3\right)\left(k^2k_1-4\kappa^{0123}\omega k_2k_3\right)}{k^6-4(\kappa^{0123})^2k^2k_3^2(\omega^2-4k_1^2-k_2^2)-32(\kappa^{0123})^3\omega k_1k_2k_3^3}\,, \\[2ex]
\varepsilon^1&=\frac{4\kappa^{0123}\left[k^4\omega k_2-\kappa^{0123}k^2k_1k_3(\omega^2-k_2^2)+8(\kappa^{0123})^2\omega k_1^2k_2k_3^2\right]}{k^6-4(\kappa^{0123})^2k^2k_3^2(\omega^2-4k_1^2-k_2^2)-32(\kappa^{0123})^3\omega k_1k_2k_3^3}\,, \\[2ex]
\varepsilon^2&=-\frac{2\kappa^{0123}k^2(k^2\omega+4\kappa^{0123}k_1k_2k_3)(k^2k_1-4\kappa^{0123}\omega k_2k_3)}{k^2\left[k^6-4(\kappa^{0123})^2k^2k_3^2(\omega^2-4k_1^2-k_2^2)-32(\kappa^{0123})^3\omega k_1k_2k_3^3\right]}\,, \\[2ex]
\varepsilon^3&=-1\,.
\end{align}
In these expressions $k^0$ has to be replaced by $\omega_1$ to obtain the polarization vector $\varepsilon^{(1)\,\mu}$ of the first
propagation mode and by $\omega_2$, respectively, to get the the polarization vector $\varepsilon^{(2)\,\mu}$ of the second mode
(see \eqref{eq:dispersion-relations-birefringent} for $\omega_{1,2}$). The general polarization tensor coefficients will be listed
as follows where
\begin{align}
\mathcal{N}&=k^{12}+8(\kappa^{0123})^2k^8k_3^2(4k_1^2+k_2^2-\omega^2)-64(\kappa^{0123})^3k^6\omega k_1k_2k_3^3 \notag \\
&\phantom{{}={}}+16(\kappa^{0123})^4k^4k_3^4(4k_1^2+k_2^2-\omega^2)^2-256(\kappa^{0123})^5k^2\omega k_1k_2k_3^5(4k_1^2+k_2^2-\omega^2) \notag \\
&\phantom{{}={}}+1024(\kappa^{0123})^6\omega^2k_1^2k_2^2k_3^6\,,
\end{align}
appears as a denominator in (almost) all these coefficients:
\begin{subequations}
\label{eq:general-polarization-tensor-coefficient-1}
\begin{align}
\widehat{a}_i&=-\frac{4\kappa^{0123}k^2}{k_3}\frac{\alpha_1\alpha_2}{\mathcal{N}}\,, \\[2ex]
\alpha_1&=k^4\omega k_2-\kappa^{0123}k^2k_1k_3(\omega^2-k_2^2)+8(\kappa^{0123})^2\omega k_1^2k_2k_3^2\,, \\[2ex]
\alpha_2&=k^4k_1+4\kappa^{0123}k^2\omega k_2k_3-8(\kappa^{0123})^2k_1k_3^2(\omega^2-2k_1^2-k_2^2)\,.
\end{align}
\end{subequations}
\begin{subequations}
\begin{align}
\widehat{b}_i&=\frac{4\kappa^{0123}k^2}{k_1k_3}\frac{\beta}{\mathcal{N}}\,, \\[2ex]
\beta&=\,-\,k^8\omega k_2(\omega^2-k_1^2)+\kappa^{0123}k^6k_1k_3\left[\omega^4-\omega^2(k_1^2+k_2^2)+2k_1^2k_2^2\right] \notag \\
&\phantom{{}={}}+4(\kappa^{0123})^2k^4\omega k_2k_3^2\left[\omega^4+\omega^2(-4k_1^2+3k_2^2)+4k_1^4\right] \notag \\
&\phantom{{}={}}+4(\kappa^{0123})^3k^2k_1k_3^3\left[-\omega^6+2\omega^4(k_1^2-5k_2^2)+\omega^2(12k_1^2k_2^2+7k_2^4)+2k_1^2k_2^2(2k_1^2+k_2^2)\right] \notag \\
&\phantom{{}={}}+64(\kappa^{0123})^4\omega k_1^2k_2k_3^4(\omega^2-k_1^2)(\omega^2-2k_1^2-k_2^2)\,.
\end{align}
\end{subequations}
\begin{subequations}
\begin{align}
\widehat{d}_i&=-\frac{4\kappa^{0123}k^2}{k_1k_3}\frac{\delta}{\mathcal{N}}\,, \\[2ex]
\delta&=k^8\omega k_2(k_1^2+k_2^2)+\kappa^{0123}k^6k_1k_3\left[-\omega^2(2k_1^2+k_2^2)+k_2^2(k_1^2+k_2^2)\right] \notag \\
&\phantom{{}={}}+4(\kappa^{0123})^2k^4\omega k_2k_3^2\left[3\omega^2k_2^2+(2k_1^2+k_2^2)^2\right] \notag \\
&\phantom{{}={}}-4(\kappa^{0123})^3k^2k_1k_3^3\left[\omega^4(-2k_1^2+7k_2^2)+2\omega^2(2k_1^4-6k_1^2k_2^2-5k_2^4)-k_2^4(2k_1^2+k_2^2)\right] \notag \\
&\phantom{{}={}}-64(\kappa^{0123})^4\omega k_1^2k_2k_3^4(k_1^2+k_2^2)(\omega^2-2k_1^2-k_2^2)\,.
\end{align}
\end{subequations}
\begin{subequations}
\begin{align}
\widehat{e}_i&=-\frac{k^2}{k_1k_3}\frac{\epsilon_1\epsilon_2}{\mathcal{N}}\,, \\[2ex]
\epsilon_1&=k^4k_1+4\kappa^{0123}k^2\omega k_2k_3-8(\kappa^{0123})^2k_1k_3^2(\omega^2-2k_1^2-k_2^2)\,, \\[2ex]
\epsilon_2&=\,-\,k^6k_3+4\kappa^{0123}k^4\omega k_1k_2+32(\kappa^{0123})^3\omega k_1k_2k_3^2(k_1^2+k_3^2) \notag \\
&\phantom{{}={}}-4(\kappa^{0123})^2k^2k_3\left[\omega^2(k_1^2-k_3^2)-k_1^2k_2^2+4k_1^2k_3^2+k_2^2k_3^2\right]\,.
\end{align}
\end{subequations}
\begin{subequations}
\begin{align}
\widehat{g}_i&=\frac{4\kappa^{0123}k^2}{k_1k_3}\frac{\gamma}{\mathcal{N}}\,, \\[2ex]
\gamma&=\,-\,k^8\omega^2k_2^2-\kappa^{0123}k^6\omega k_1k_2k_3\left[k_1^2+3(k_2^2-\omega^2)\right] \notag \\
&\phantom{{}={}}-2(\kappa^{0123})^2k^4k_3^2(\omega^2-k_2^2)\left[\omega^2(k_1^2+2k_2^2)-k_1^2(2k_1^2+k_2^2)\right] \notag \\
&\phantom{{}={}}+4(\kappa^{0123})^3k^2\omega k_1k_2k_3^3\left[\omega^4-2\omega^2(2k_1^2+7k_2^2)+(2k_1^2+k_2^2)^2\right] \notag \\
&\phantom{{}={}}+64(\kappa^{0123})^4\omega^2k_1^2k_2^2k_3^4(\omega^2-2k_1^2-k_2^2)\,.
\end{align}
\end{subequations}
\begin{subequations}
\begin{align}
\widehat{h}_i&=-\frac{2\kappa^{0123}k^2}{k_1}\frac{\theta_1\theta_2\theta_3}{\mathcal{N}}\,, \\[2ex]
\theta_1&=-k^2k_2+4\kappa^{0123}\omega k_1k_3\,, \\[2ex]
\theta_2&=-k^2k_1+4\kappa^{0123}\omega k_2k_3\,, \\[2ex]
\theta_3&=k^4k_1+4\kappa^{0123}k^2\omega k_2k_3-8(\kappa^{0123})^2k_1k_3^2(\omega^2-2k_1^2-k_2^2)\,.
\end{align}
\end{subequations}
\begin{subequations}
\begin{align}
\widehat{k}_i&=\frac{2\kappa^{0123}k^2}{k_1}\frac{\kappa_1\kappa_2\kappa_3}{\mathcal{N}}\,, \\[2ex]
\kappa_1&=\omega^3-\omega\mathbf{k}^2+4\kappa^{0123}k_1k_2k_3\,, \\[2ex]
\kappa_2&=k^2k_1-4\kappa^{0123}\omega k_2k_3\,, \\[2ex]
\kappa_3&=k^4k_1+4\kappa^{0123}k^2\omega k_2k_3-8(\kappa^{0123})^2k_1k_3^2(\omega^2-2k_1^2-k_2^2)\,.
\end{align}
\end{subequations}
\begin{subequations}
\begin{align}
\widehat{l}_i&=\frac{4\kappa^{0123}}{k_1k_3}\frac{\lambda_1}{\lambda_2}\,, \\[2ex]
\lambda_1&=k^4\omega k_2-\kappa^{0123}k^2k_1k_3(\omega^2-k_2^2)+8(\kappa^{0123})^2\omega k_1^2k_2k_3^2\,, \\[2ex]
\lambda_2&=-k^6+4(\kappa^{0123})^2k^2k_3^2(\omega^2-4k_1^2-k_2^2)+32(\kappa^{0123})^3\omega k_1k_2k_3^3\,.
\end{align}
\end{subequations}
\begin{subequations}
\begin{align}
\widehat{m}_i&=\frac{4\kappa^{0123}k^2}{k_1k_3}\frac{\mu_1\mu_2}{\mathcal{N}}\,, \\[2ex]
\mu_1&=k^4\omega k_2-\kappa^{0123}k^2k_1k_3(\omega^2-k_2^2)+8(\kappa^{0123})^2\omega k_1^2k_2k_3^2\,, \\[2ex]
\mu_2&=k^4\omega+2\kappa^{0123}k^2k_1k_2k_3-4(\kappa^{0123})^2\omega k_3^2(\omega^2-2k_1^2+k_2^2)\,.
\end{align}
\end{subequations}
\begin{subequations}
\label{eq:general-polarization-tensor-coefficient-10}
\begin{align}
\widehat{o}_i&=\frac{4\kappa^{0123}k^2}{k_1k_3}\frac{\rho_1\rho_2}{\mathcal{N}}\,, \\[2ex]
\rho_1&=k^4\omega k_2-\kappa^{0123}k^2k_1k_3(\omega^2-k_2^2)+8(\kappa^{0123})^2\omega k_1^2k_2k_3^2\,, \\[2ex]
\rho_2&=k^4k_2-2\kappa^{0123}k^2\omega k_1k_3+4(\kappa^{0123})^2k_2k_3^2(\omega^2+2k_1^2+k_2^2)\,.
\end{align}
\end{subequations}
The index $i=1$, 2 indicates the coefficients of the first and the second propagation mode, respectively. The normalization of the
polarization vectors can be cast in the following form:
\begin{align}
\label{eq:general-polarization-vector-normalization}
N&=\frac{k^4}{2\omega^2}\left(8\kappa^{0123}k^6\omega k_1k_2k_3+k^8[\omega^2+k_1^2+k_2^2]\right. \notag \\
&\phantom{{}={}2\omega^2\Big\{}\left.+\,4(\kappa^{0123})^2k^4\left\{\omega^4(k_1^2+4k_2^2-2k_3^2)+\omega^2\left[k_1^4+4k_2^2(k_2^2+k_3^2)+k_1^2(2k_2^2+k_3^2)\right]\right.\right. \notag \\
&\phantom{{}={}2\omega^2\Big\{}\left.\phantom{+\,4(\kappa^{0123})^2k^4\Big[}+k_1^4(k_2^2+8k_3^2)+k_1^2k_2^2(k_2^2+9k_3^2)+2k_2^4k_3^2\big\}\right. \notag \\
&\phantom{{}={}2\omega^2\Big\{}\left.+\,32(\kappa^{0123})^3k^2\omega k_1k_2k_3\left[-2\omega^4+\omega^2(k_1^2-2k_2^2-3k_3^2)+(k_1^2k_2^2+4k_1^2k_3^2+k_2^2k_3^2)\right]\right. \notag \\
&\phantom{{}={}2\omega^2\Big\{}\left.+\,16(\kappa^{0123})^4k_3^2\left\{4k_1^6(\omega^2+k_2^2+4k_3^2)+4k_1^4\left[k_2^4+5k_2^2k_3^2-\omega^2(\omega^2+3k_3^2)\right]\right.\right. \notag \\
&\phantom{{}={}2\omega^2\Big\{+16(\kappa^{0123})^4k_3^2\Big[}\left.+\,k_1^2\left[\omega^6+\omega^2k_2^2(3\omega^2-8k_3^2)+k_2^4(3\omega^2+8k_3^2)+k_2^6\right]\right. \notag \\
&\phantom{{}={}2\omega^2\Big\{+16(\kappa^{0123})^4k_3^2\Big[+\,k_1^2\Big[}\left.\left.+\,(\omega^2+k_2^2)^2\left[\omega^2k_3^2+k_2^2(4\omega^2+k_3^2)\right]\right\}\right)\mathcal{N}^{-1}\,.
\end{align}
For each mode $k^0$ is again understood to be replaced by $\omega_1$ and $\omega_2$, respectively, where $N'\equiv N|_{k^0=\omega_1}$
and $N''\equiv N|_{k^0=\omega_2}$. The coefficients of the general propagator in Feynman gauge are multiplied by the denominator
\begin{align}
\widehat{K}^{-1}&=k^4-4(\kappa^{0123})^2\left[k_0^2(k_1^2+4k_2^2+k_3^2)-(k_1^2k_2^2+4k_1^2k_3^2+k_2^2k_3^2)\right] \notag \\
&\phantom{{}={}}+32(\kappa^{0123})^3k_0 k_1k_2k_3\,,
\end{align}
and they are given by:
\begin{subequations}
\label{eq:general-propagator-coefficient-1}
\begin{align}
\widehat{a}&=\frac{1}{k^2k_3}\widetilde{\alpha}\,, \\[2ex]
\widetilde{\alpha}&=k^4k_3+4\kappa^{0123}k^2k_0 k_1k_2-8(\kappa^{0123})^2k_1^2k_3(k_0^2-k_2^2-2k_3^2)\,.
\end{align}
\end{subequations}
\begin{subequations}
\begin{align}
\widehat{b}&=-\frac{4\kappa^{0123}}{k^2k_1k_3}\widetilde{\beta}\,, \\[2ex]
\widetilde{\beta}&=-k^2k_0 k_2(k_0^2-k_1^2-k_3^2)+\kappa^{0123}k_1k_3\left[k_0^4-k_0^2(k_1^2+k_2^2+3k_3^2)+2k_1^2(k_2^2+2k_3^2)\right]\,.
\end{align}
\end{subequations}
\begin{subequations}
\begin{align}
\widehat{d}&=-\frac{4\kappa^{0123}}{k^2k_1k_3}\widetilde{\delta}\,, \\[2ex]
\widetilde{\delta}&=-k^2\mathbf{k}^2k_0 k_2+\kappa^{0123}k_1k_3\left[k_0^2(2k_1^2+k_2^2)-k_1^2(k_2^2+4k_3^2)-k_2^2(k_2^2+3k_3^2)\right]\,.
\end{align}
\end{subequations}
\begin{subequations}
\begin{align}
\widehat{e}&=\frac{4\kappa^{0123}(k_3^2-k_1^2)}{k^2k_1k_3}\widetilde{\epsilon}\,, \\[2ex]
\widetilde{\epsilon}&=-k^2k_0 k_2+2\kappa^{0123}k_1k_3(k_0^2-k_2^2)\,.
\end{align}
\end{subequations}
\begin{subequations}
\begin{align}
\widehat{g}&=\frac{2\kappa^{0123}}{k^2k_1k_3}\widetilde{\gamma}\,, \\[2ex]
\widetilde{\gamma}&=k^2\left[k_0^2(2k_2^2+k_3^2)-k_3^2(2k_1^2+k_2^2)\right]-2\kappa^{0123}k_0 k_1k_2k_3(3k_0^2-k_1^2-3k_2^2-5k_3^2)\,.
\end{align}
\end{subequations}
\begin{subequations}
\begin{align}
\widehat{h}&=\frac{2\kappa^{0123}(k_3^2-k_1^2)}{k^2k_1}\widetilde{\theta}\,, \\[2ex]
\widetilde{\theta}&=-k^2k_2+4\kappa^{0123}k_0 k_1k_3\,.
\end{align}
\end{subequations}
\begin{subequations}
\begin{align}
\widehat{k}&=\frac{2\kappa^{0123}(k_1^2-k_3^2)}{k^2k_1}\widetilde{\kappa}\,, \\[2ex]
\widetilde{\kappa}&=-k^2k_0-4\kappa^{0123}k_1k_2k_3\,.
\end{align}
\end{subequations}
\begin{subequations}
\begin{align}
\widehat{l}&=\frac{4\kappa^{0123}}{k^4k_1k_3}\widetilde{\lambda}\,, \\[2ex]
\widetilde{\lambda}&=k^4k_0 k_2-\kappa^{0123}k^2k_1k_3(k_0^2-k_2^2)+8(\kappa^{0123})^2k_0 k_1^2k_2k_3^2\,.
\end{align}
\end{subequations}
\begin{subequations}
\begin{align}
\widehat{m}&=-\frac{2\kappa^{0123}}{k^2k_1k_3}\widetilde{\mu}\,, \\[2ex]
\widetilde{\mu}&=k^2k_2(2k_0^2-k_3^2)-2\kappa^{0123}k_0 k_1k_3(k_0^2-3k_2^2-2k_3^2)\,.
\end{align}
\end{subequations}
\begin{subequations}
\label{eq:general-propagator-coefficient-10}
\begin{align}
\widehat{o}&=\frac{2\kappa^{0123}}{k^2k_1k_3}\widetilde{\rho}\,, \\[2ex]
\widetilde{\rho}&=-k^2k_0(2k_2^2+k_3^2)+2\kappa^{0123}k_1k_2k_3(3k_0^2-k_2^2-2k_3^2)\,.
\end{align}
\end{subequations}
All polarization tensor coefficients and propagator coefficients that are not listed vanish.
For a set of randomly chosen momentum components and Lorentz-violating parameter $\kappa^{0123}$ such as $\mathbf{k}=(3,5,7)/\mathrm{m}$
plus $\kappa^{0123}=2$ the following relationships can be shown numerically for the results above:
\begin{subequations}
\begin{align}
\frac{\{\widehat{a}_1,\widehat{b}_1,\widehat{d}_1,\widehat{e}_1,\widehat{g}_1,\widehat{h}_1,\widehat{k}_1,\widehat{l}_1,\widehat{m}_1,\widehat{o}_1\}}{N'}&=-2\omega \left.\frac{\{\widehat{a},\widehat{b},\widehat{d},\widehat{e},\widehat{g},\widehat{h},\widehat{k},\widehat{l},\widehat{m},\widehat{o}\}}{(\omega-\omega_2)(\omega-\widetilde{\omega}_1)(\omega-\widetilde{\omega}_2)}\right|_{\substack{k^0=\omega_1 \\ \omega=\omega_1}}\,, \\[2ex]
\frac{\{\widehat{a}_2,\widehat{b}_2,\widehat{d}_2,\widehat{e}_2,\widehat{g}_2,\widehat{h}_2,\widehat{k}_2,\widehat{l}_2,\widehat{m}_2,\widehat{o}_2\}}{N''}&=-2\omega \left.\frac{\{\widehat{a},\widehat{b},\widehat{d},\widehat{e},\widehat{g},\widehat{h},\widehat{k},\widehat{l},\widehat{m},\widehat{o}\}}{(\omega-\omega_1)(\omega-\widetilde{\omega}_1)(\omega-\widetilde{\omega}_2)}\right|_{\substack{k^0=\omega_2 \\ \omega=\omega_2}}\,,
\end{align}
\end{subequations}
where $\widetilde{\omega}_{1,2}$ correspond to the negative counterparts of the physical propagator poles.
According to the lines of \secref{sec:discussion-of-unitarity}, the optical theorem is then checked to be valid numerically. Two remarks are
in order. First, the propagator with the coefficients of Eqs.~(\ref{eq:general-propagator-coefficient-1}) -- (\ref{eq:general-propagator-coefficient-10})
reduces to the standard one in Feynman gauge for the limit $\kappa^{0123}\mapsto 0$. Second, only two out of
three preferred spacelike vectors of \eqref{eq:orthonormal-set-preferred-vectors} are needed to write the propagator plus the polarization
tensors covariantly. If the vector $\xi_1=(0,1,0,0)^T$ is not used, which was done here, the respective expressions are much simpler. This is
the reason why for $k_1\mapsto 0$ and $k_2\mapsto 0$ the results of \secref{sec:propagator-polarization-vectors} are not recovered since in the
latter section the background vector $\xi_1$ is, indeed, employed.

\end{appendix}

\newpage


\end{document}